\documentstyle[11pt,aaspp4,flushrt,psfig]{article}  
\received{RECEIPT DATE 1996}
\accepted{ACCEPT DATE 1996}
\journalid{VOL}{JOURNAL DATE 1996}
\articleid{START PAGE}{END PAGE}
\font\fiverm=cmr5

          \font\sixrm=cmr6       
                     
\def\app{Astroparticle Phys.}                   
\def\asr{Adv. Space Res.}                       
\def\ssr{Space Sci. Rev.}                       
\def\teq#1{$\, #1\,$}                     
\def\kB{k_{\hbox{\fiverm B}}}       
\def\dover#1#2{\hbox{${{\displaystyle#1 \vphantom{(} }\over{
   \displaystyle #2 \vphantom{(} }}$}}
\def\rg{r_{\rm g}}
\def\vf{v_{\hbox{\sixrm F}}}
\def\vxf{v_{\rm x\hbox{\sixrm F}}}
\def\vht{v_{\hbox{\sixrm HT}}}
\def\vxsk{v_{\rm xsk}}
\def\Tbn{\Theta_{\rm Bn}}
\def\Tbnone{\Theta_{\rm Bn1}}
\def\Tbntwo{\Theta_{\rm Bn2}}
\def\PitchB{\theta_{\hbox{\sixrm B}}}
\def\PitchBnew{\theta_{\hbox{\sixrm B}}^{\hbox{\sixrm N}}}
\def\PitchBold{\theta_{\hbox{\sixrm B}}^{\hbox{\sixrm O}}}
\def\PhaseB{\phi_{\hbox{\sixrm B}}}
\def\phiBnew{\phi_{\hbox{\sixrm B}}^{\hbox{\sixrm N}}}
\def\phiBold{\phi_{\hbox{\sixrm B}}^{\hbox{\sixrm O}}}
\def\dFEB{d_{\hbox{\sixrm FEB}}}

\def\vA{v_{\hbox{\sixrm A}1}}
\def\MA{M_{\hbox{\sixrm A}1}}
\def\MS{M_{\hbox{\sixrm S}1}}
\def\Fuxb{F_{\rm uB}^{'}}
\def\FenUpS{F_{\rm en1}^{'}}
\def\Fen{F_{\rm en}^{'}(x)}
\def\FpxxUpS{F_{\rm xx1}^{'}}
\def\Fpxx{F_{\rm xx}^{'}(x)}
\def\Fpxz{F_{\rm xz1}^{'}}
\begin{document}
%
%
\newcommand{\vol}[2]{$\;$ \bf #1\rm , #2}                 
\newcommand{\figureout}[2]{ \figcaption[#1]{#2} }       
%
%
%
\title{NON-LINEAR PARTICLE ACCELERATION IN OBLIQUE SHOCKS}
   \author{Donald C. Ellison}                      
   \affil{Department of Physics, North Carolina State University, \\
      Box 8202, Raleigh NC 27695, U.S.A.\\
      don\_ellison@ncsu.edu}
   \author{Matthew G. Baring\altaffilmark{1} and Frank C. Jones}
   \altaffiltext{1}{Compton Fellow, Universities Space Research Association} 
   \affil{Laboratory for High Energy Astrophysics, \\
      NASA Goddard Space Flight Center, Greenbelt, MD 20771, U.S.A.\\
       baring@lheavx.gsfc.nasa.gov, frank.c.jones@gsfc.nasa.gov}
   \authoraddr{Department of Physics, North Carolina State University,
       Box 8202, Raleigh NC 27695, U.S.A.}
\begin{abstract}
The solution of the non-linear diffusive shock acceleration problem,
where the pressure of the non-thermal population is sufficient to
modify the shock hydrodynamics, is widely recognized as a key to
understanding particle acceleration in a variety of astrophysical
environments.  We have developed a Monte Carlo technique for
self-consistently calculating the hydrodynamic structure of oblique,
steady-state shocks, together with the first-order Fermi acceleration
process and associated non-thermal particle distributions.  This is
the first internally consistent treatment of modified shocks that
includes cross-field diffusion of particles.  Our method overcomes the
injection problem faced by analytic descriptions of shock
acceleration, and the lack of adequate dynamic range and artificial
suppression of cross-field diffusion faced by plasma simulations; it
currently provides the most broad and versatile description of
collisionless shocks undergoing efficient particle acceleration.  We
present solutions for plasma quantities and particle distributions
upstream and downstream of shocks, illustrating the strong differences
observed between non-linear and test-particle cases. It is found that
there are only marginal differences in the injection efficiency and
resultant spectra for two extreme scattering modes, namely large-angle
scattering and pitch-angle diffusion, for a wide range of shock
parameters, i.e., for non-perpendicular subluminal shocks with field
obliquities less than or equal to \teq{75^\circ} and de Hoffmann-Teller
frame speeds much less than the speed of light.  
\end{abstract}
\clearpage

\section{INTRODUCTION}

The importance of shocks as generators of highly non-thermal particle
distributions in heliospheric and astrophysical environments has been
well-documented in the literature (e.g. Axford 1981; V\"olk 1984; Blandford and
Eichler 1987; Jones and Ellison 1991).  While direct detections of high energy
particles are obtained via terrestrial observations of the cosmic ray flux and
spacecraft measurements of non-thermal ions in the solar neighbourhood and in
environs of planetary bow shocks and interplanetary travelling shocks, the
existence of abundant non-thermal particle populations in a diversity of
astrophysical locales can be inferred from the prominence of non-thermal
radiation emitted by many cosmic objects.
Understanding the details of shock acceleration is of critical importance since
many such objects emit predominantly non-thermal radiation, and indeed some
sources are observed \it only \rm because they produce non-thermal particles
(e.g. radio emission from supernova remnants and extra-galactic jets).  The
first-order Fermi mechanism of diffusive shock acceleration is the most popular
candidate for particle energization at astrophysical shocks. The test-particle
(i.e. linear) theory (Krymsky 1977; Axford et al. 1977; Bell 1978; Blandford
and Ostriker 1978) of this process is straightforward and yields the most
important result, namely that a power-law with a spectral index that is
relatively insensitive to the ambient conditions, is the natural product of
collisionless shock acceleration.

The equally important question of the efficiency of the process can only be
adequately addressed with a fully non-linear (and therefore complex)
calculation. The inherent efficiency of shock acceleration, which is evident in
observations at the Earth's bow shock (e.g. Ellison, M\"obius, and Paschmann
1990), and in modelling of plane-parallel (e.g. see Ellison and Eichler 1985;
Ellison, Jones and Reynolds 1990) shocks, where the field is normal to the
shock, and oblique shocks (e.g. Baring, Ellison and Jones 1993; Ellison, Baring
and Jones 1995), implies that hydrodynamic feedback effects between the
accelerated particles and the shock structure are very important and therefore
essential to any complete description of the process. This has turned out to be
a formidable task because of the wide range of spatial and energy scales that
must be self-consistently included in a complete calculation.  On the one hand,
the microphysical plasma processes of the shock dissipation control injection
from the thermal population and on the other hand, the highest energy particles
(extending to at least \teq{10^{14}}eV in the case of galactic cosmic rays)
with extremely long diffusion lengths, are dynamically significant in strong
shocks and feedback on the shock structure. Ranges of interacting scales of
many orders of magnitude must be described self-consistently.

Additional complications stem from the fact that the geometry of shocks, i.e.,
whether they are  oblique or parallel, strongly affects the acceleration
efficiency (e.g., Ellison, Baring, and Jones 1995), even though the
test-particle result is independent of the geometry.  Observations indicate
that interplanetary shocks, bow shocks (both planetary and from jets), the
solar wind termination shock, and supernova remnant blast waves have a wide
range of obliquities thereby rendering considerations of shock geometry
salient. It turns out that the angle between the upstream magnetic field and
the shock normal, \teq{\Tbnone}, is a decisive parameter in determining all
aspects of the shock, including the ability to inject and accelerate particles,
and therefore has obvious observational consequences.  For instance, diffuse
ions generated at quasi-parallel portions of the Earth's bow shock differ
radically in energy content, distribution function, etc., from field aligned
beams generated at quasi-perpendicular portions of the shock (e.g. Ipavich et
al.  1988).  Furthermore, the observed variation of radio intensity around the
rim of shell-like supernova remnants may be the result of varying shock
obliquity (e.g. Fulbright and Reynolds 1990), and the acceleration of the
anomalous cosmic ray component at the solar wind termination shock may depend
on rapid acceleration rates obtained in highly oblique portions of the shock
(Jokipii 1992).  Unfortunately, in models which ignore the plasma
microstructure as we do here, oblique shocks are more complicated and require
additional parameters for a complete description than do parallel
(i.e. $\Tbnone=0^\circ$) ones, primarily the degree of diffusion perpendicular
to the mean ambient magnetic field direction.

In this paper we present our method for calculating the structure of
steady-state, collisionless shocks of arbitrary obliquity and with efficient
particle injection and acceleration.  The method, a computer simulation using
Monte Carlo techniques, is an extension of our previous work on modified
parallel shocks (e.g. see Jones and Ellison 1991 and references therein), where
we explored the properties of the non-linear modified shock scenario, and
test-particle oblique shocks (Ellison, Baring, and Jones 1995; Baring, Ellison,
and Jones 1993, 1995), where we determined the dependence of acceleration
efficiency on obliquity \teq{\Tbnone}.  These studies have been successfully
applied to AMPTE observations near the parallel portion of the Earth's bow
shock (Ellison, M\"obius, and Paschmann 1990), a high Mach number shock with
strong modification by the accelerated ions, and measurements by Ulysses at
highly oblique travelling interplanetary shocks in the heliosphere (Baring et
al. 1995), which generally have low Mach numbers and therefore are
well-modelled by linear test-particle simulations.  The impressive fits
obtained to the spectral data (i.e. ion distribution functions) from each of
these experiments underlines the importance of the Fermi mechanism and the 
value of the Monte Carlo technique.  The present work represents the first
self-consistent treatment of modified shocks that includes three-dimensional
diffusion.

With the Monte Carlo simulation, we self-consistently determine the average
flow speed and magnetic field structure across the shock under the influence of
accelerated particles maintaining constant particle, momentum, and energy
fluxes at all positions from far upstream to far downstream of the shock. 
Particles are injected upstream of the shock, propagated and diffused in the
shock environs until they eventually leave the system.  We calculate their
orbits exactly as in the works of Decker (1988), Begelman and Kirk (1990),
Ostrowski (1991) and our recent test-particle treatment (Ellison, Baring and
Jones 1995), and make no assumption relating to the particle magnetic moment. 
Our method does not self-consistently calculate the complex plasma processes
responsible for dissipation, but instead postulates that these processes can be
adequately described with a simple elastic scattering relation that is assumed
to be valid for \it all \rm particle energies; thermal and non-thermal
particles are treated identically.  This simplification sacrifices the details
of wave-particle interactions, but permits simultaneous description of the
thermal plasma and the particle injection and acceleration to the high energies
associated with space plasma shocks, thereby satisfying the aforementioned goal
of broad dynamic range. Cross-field diffusion is included via a parametric
description but is fully three-dimensional in contrast to hybrid or full plasma
simulations with one or two ignorable dimensions which suffer from artificial
suppression of  cross-field diffusion (e.g. Jokipii, Giacalone, and K\'ota
1993). Simulation output includes the ion distribution function at all relevant
positions in the shocked flow, for a range of obliquities and Mach numbers. 

Results are compared (in Section~4) for two extreme scattering modes, namely
large-angle scattering (LAS), where the direction of a particle is isotropized
in a single scattering event, and pitch-angle diffusion (PAD), where small
changes in the angle a particle's momentum makes with the local magnetic field
occur at each time step. The former of these extremes mimics particle motion in
highly turbulent fields, while the latter is usually implemented in analytic
treatments of Fermi acceleration (e.g. Kirk and Schneider 1987, but see also
the Monte Carlo work of Ostrowski 1991). We find that in our application to
non-relativistic shocks, the choice of scattering mode is largely immaterial to
the resultant distributions; we expect this not to be so for relativistic
shocks where the modes generate vastly different particle anisotropies. We also
compare non-linear (Section~4.2) results with test-particle ones where the
non-thermal particles do not modify a discontinuous shock (Section~4.1),
finding that, as in our earlier work on plane-parallel shocks, some spectral
curvature arises in high Mach number shocks where a large fraction of the
partial pressure resides in the non-thermal population.  An outline of the
Monte Carlo method is given in Section~2, followed in Section~3 by flux
conservation considerations and the associated scheme for iterative
determination of the modified shock flow and field profiles.  The spectral and
flux results comprise Section~4, culminating in a presentation of acceleration
efficiencies and discussion of the results.

\section{THE MONTE CARLO METHOD}

The Monte Carlo technique for describing  particle acceleration at  plane
shocks has been described in previous papers (e.g. Ellison, Jones and Reynolds
1990; Jones and Ellison 1991; Baring, Ellison and Jones 1993; Ellison, Baring
and Jones 1995), and is essentially a kinematic model that closely follows
Bell's (1978) approach to diffusive acceleration.  The simulation follows
individual particles as they traverse a background ``plasma'' consisting of an
average bulk flow and magnetic field; the flow velocity and magnetic field
consist of a grid of values from far upstream to far downstream with a subshock
positioned at \teq{x=0}.  Our subshock is a sharp discontinuous substructure of
the overall shock, defining the conventional boundary between (infinite)
upstream and downstream regions.  Strictly it should be no sharper than the
smallest diffusion scale (i.e. the gyroradius of thermal particles), however we
require it to be abrupt for the purposes of expedience in the simulation. 
Particles are injected far upstream of the shock with a thermal distribution at
temperature \teq{T_1}, mimicking for example solar wind ions (as in
applications to the Earth's bow shock), and are allowed to convect in the flow
and scatter, crossing the shock a few or many times before they eventually
leave the system either far downstream or beyond an upstream free escape
boundary (FEB).  They are moved one at a time according to a prescribed
scattering law, defined below, in a test-particle fashion until they exit the
simulation.  In cases where efficient acceleration arises, feedback of the
accelerated population leads to significant smoothing of the shock profile and
heating of particles occurs in the foreshock region; discussion of such
non-linear aspects of the simulation are deferred to Section~3 below.

Following our previous treatments of oblique shocks (Baring, Ellison and Jones
1993; Ellison, Baring and Jones 1995) particle convection is performed in the
de Hoffmann-Teller (HT) frame (de Hoffmann and Teller 1950), a frame where the
shock is stationary, the fluid flow \teq{{\bf u}} is everywhere parallel to the
local field \teq{{\bf B}}, and the electric field is \teq{{\bf u} \times {\bf
B}={\bf 0}} everywhere.  The HT frame of reference is therefore particularly
convenient because of the associated absence of drift electric fields: particle
trajectories are then simple gyrohelices and the description of convection is
elementary.  Furthermore, it follows from the mere existence of an HT frame
that the so-called shock drift mechanism is inseparable from, and intrinsically
part of, the Fermi acceleration process (e.g. Drury 1983; Jones and Ellison,
1991) and is therefore automatically included in our Monte Carlo technique
since particle motion is followed in the HT frame.

While field and flow directions in the HT frame are uniquely defined downstream
of the shock, the non-linear nature of this work yields a spatial variation of
\teq{{\bf u}} and \teq{{\bf B}}, due to the compressive effects of the
accelerated population.  This variation is accommodated using a grid-zone
structure that was implemented in many earlier versions of the Monte Carlo
technique; each zone contains uniform field and flow, with discontinuities at
the boundaries satisfying the Rankine-Hugoniot conditions discussed in
Section~3 below.  The grid-zone boundaries therefore mimic mini-shocks, with
the subshock defining a particular grid boundary; a depiction of this grid
structure is given in Baring, Ellison and Jones (1992). The spatial resolution
of the grid can be adapted at will, but in our applications, we require it to
be finer at some distance upstream than the typical mean free path of particles
that penetrate to that distance from the shock.  Note that the ability to
define (for example via the Rankine-Hugoniot conditions) an HT frame with
\teq{{\bf u} \times {\bf B}={\bf 0}} on both sides of a grid point implies, by
spatial extension, that the HT frame is uniquely defined throughout the flow,
regardless of flow and field compression and deflection upstream. In addition,
note that even though the \teq{{\bf u} \times {\bf B}} electric field is
transformed away by going to the HT frame, charge separation electric fields
are not and these have not been included in our model, being beyond the scope
of the present work.

While particle transport is monitored in the HT frame, all measured quantities
such as particle distributions and momentum and energy fluxes are output in the
normal incidence frame (NIF), which is the frame where the shock is also
stationary, but is defined such that the flow far upstream (i.e. where it is
uniform to infinity) is normal to the shock plane.  A simple velocity boost
with a speed of \teq{\vht = u_1\tan\Tbnone} parallel to the shock front effects
transformation between the NIF and HT frames, where \teq{\Tbnone} is the far
upstream angle the magnetic field makes with the shock normal and \teq{u_1} is
the far upstream flow speed in the frame where the shock is at rest. Note that
hereafter, the index `1' will indicate far upstream values and the index `2'
will indicate far downstream values well away from the smooth shock transition. 
A depiction of the NIF geometry is given in Figure~\ref{fig:shock} for the
specific case of unmodified shocks; modified shock geometry extends this to
include piecewise increments of \teq{u} and \teq{B}.  In the normal incidence
frame, the $-x$-axis defines the shock normal, and the senses of the other axes
are as in Figure~\ref{fig:shock}.  The results of this paper are restricted to
highly subluminal cases where \teq{\vht\ll c}, the speed of light, since our
Monte Carlo technique has not yet been generalized to include relativistic
effects in oblique shocks.

\placefigure{fig:shock}          

The simulation is an orbit code, where particle propagation is performed by
following the gyromotions exactly, as in the test-particle work of Ellison,
Baring and Jones (1995) and a diversity of works in the literature (e.g. Decker
1988; Begelman and Kirk 1990; Ostrowski 1991; Takahara and Terasawa 1991). The
position of particles is incrementally updated on a timescale \teq{\delta t},
which is a small fraction of a gyroperiod, i.e., \teq{\delta t\ll\tau_{\rm
g}=mc/(QeB)}, where \teq{m} and \teq{Q} are the particle's mass and charge
number respectively, and \teq{e} is the electronic charge.  A particle in a
particular grid-zone is moved in a helical orbit determined by the magnetic
field and bulk flow velocity for that grid position.

\subsection{Particle Scattering}

Having outlined the procedure for convection, here we describe our prescription
for particle scattering, which is somewhat more involved. After each time step
\teq{\delta t} a determination of whether the particle should ``scatter'' or
not is made using a scattering probability \teq{P_{\rm scat} = \delta t/t_{\rm
c}} where the collision time, \teq{t_{\rm c}}, is given by \teq{\lambda /v},
\teq{\lambda} is the mean free path, and \teq{v\ll c} is the particle speed,
both measured in the local fluid frame.  This prescription yields an exponential
pathlength distribution.  The scatterings, presumably off magnetic
irregularities in the flow, are assumed to be elastic in the local plasma frame
so that monotonic energy gains  naturally arise as a particle diffuses back and
forth across the shock due to the converging nature of the flow. It proves
convenient to  scale the mean free path by the gyroradius, introducing a model
parameter \teq{\eta} that is the ratio of the two quantities (following Jokipii
1987; Ellison, Baring and Jones 1995):
\begin{equation}
   \lambda = \eta \rg \quad {\rm or} \quad
   \kappa_{\parallel} = \dover{1}{3} \eta \rg v \ ,
 \label{eq:mfp}
\end{equation}
where \teq{\kappa_{\parallel}} is the diffusion coefficient parallel to the
ambient local field.  It follows that the collision time satisfies
\begin{equation}
   t_{\rm c}\; = \;\dover{\lambda}{v}\; =\; \dover{\eta \rg}{v}\; =\; 
   \eta\,\dover{m c}{Q e B} \ .
 \label{eq:collisiontime}
\end{equation}
Generally, \teq{\eta} is a function of energy, however in this paper
it is assumed to be a constant independent of position and energy
(also following Jokipii 1987).  For this choice, the collision time is
independent of particle energy, a convenient simplification that can
be easily generalized to some other dependence of \teq{\lambda} on
\teq{\rg} (see Ellison, M\"obius, and Paschmann 1990).  Note also that
\teq{\lambda} is hence implicitly inversely proportional to \teq{B}.

The assumed constancy of \teq{\eta} is the most important approximation we
make, since all of the complicated plasma physics of wave-particle interactions
is incorporated in equation~(\ref{eq:mfp}). Convenience aside, there are sound
reasons for choosing this simple relation.  First of all, as long as
electrostatic effects are neglected (they are omitted from our treatment), the
gyroradius is the fundamental scale length of a particle at a particular energy
and the mean free path can be expected to be some function of this parameter.
Second, if the plasma is strongly turbulent with \teq{\delta B/B \sim 1}, as is
generally observed in space plasmas, the large-scale structures in the magnetic
field will mirror particles effectively on gyroradii scales (i.e. for
\teq{\eta\approx 1}, the Bohm diffusion limit; Zachary 1987).   Third, and most
important, spacecraft observations suggest that \teq{\lambda \propto \rg} in
the self-generated turbulence near the earth's bow shock (e.g. Ellison,
M\"obius, and Paschmann 1990).  Fourth, hybrid plasma simulation results also
suggest that the mean free path is a moderately increasing function of
\teq{\rg} (Giacalone, Burgess, and Schwartz 1992; Ellison et al. 1993). Note
that while the exact form for \teq{\lambda} will surely have quantitative
effects on the injection rate and all other shock characteristics, the most
important qualitative effects should be well modeled as long as a strongly
energy dependent diffusion coefficient is used.  Employing a realistic
\teq{\kappa_{\parallel}} with a strong  energy dependence is essential because
of the intrinsic efficiency of shock acceleration.  If \teq{\kappa_{\parallel}}
is indeed energy dependent, the highest energy particles with large fractions
of energy and pressure have very different scales from thermal particles,
leading to the spectral curvature that appears in the simulated distributions
(see Section~4).

The simulation employs two complementary types of scattering modes, namely
large-angle scattering and pitch-angle diffusion.  For each mode, elasticity of
scattering is imposed, which amounts to neglecting any recoil effects of wave
production on the particles and hence that the background scattering centers
(i.e. magnetic irregularities) are frozen in the plasma.  This approximation is
generally quite appropriate but becomes less accurate for low Afv\'enic Mach
numbers when the flow speed does not far exceed the Alfv\'en wave speed.  In
large-angle scattering (LAS), a particle's direction is randomized in a single
scattering event (on a timescale of \teq{t_{\rm c}}) and the new direction is
made isotropic in the local plasma frame.  Such quasi-isotropic scattering is
adopted in most of our earlier simulation work (e.g. see Jones and Ellison
1991; Baring, Ellison and Jones 1993; Ellison Baring and Jones 1995), and is
intended to mimic the effect of large amplitude field turbulence on particle
motions. Such turbulence is present in both plasma simulations (e.g. Quest
1988; Burgess 1989; Winske et al. 1990) and observations of shocks in the
heliosphere (e.g., Hoppe et al. 1981).

The second scattering mode we employ is pitch-angle diffusion (PAD), as used in
the diverse works of Decker and Vlahos (1985), Decker (1988), Kirk and
Schneider (1987) and Ostrowski (1988, 1991).  In this mode, the direction of
the velocity vector \teq{{\bf v}} is changed by a small amount after each time
step, \teq{\delta t}, rendering the scattering process more ``continuous'' than
large angle collisions, and more appropriate to physical systems with small
levels of field turbulence.  Here we adopt the procedure detailed in Ellison,
Jones, and Reynolds (1990) for determining the maximum amount the pitch angle
\teq{\PitchB = \arccos\{ {\bf v}\cdot {\bf B}/ \vert {\bf v}\vert\,\vert{\bf
B}\vert\}} can change after each \teq{\delta t}; our procedure is summarized in
the Appendix.  A comparison of the simulation results for these two modes is
one goal of this paper, motivated by an expectation that they could, in
principal, produce different injection efficiencies at modified shocks.  This
expectation is partly based on the spectral differences  observed between LAS
and PAD applications to unmodified relativistic shocks (e.g. Ellison, Jones,
and Reynolds 1990), where the two modes generate significantly different
particle anisotropies; distribution anisotropies are indeed relevant to the
injection problem considered here.

As a particle convects, the simulation tracks both its position and the
position of its gyrocenter.  After a scattering occurs and a new direction is
obtained for its velocity vector, a new gyrocenter is calculated.  This shift
of the gyrocenter means the particle is now gyrating around a different field
line and diffusion across the field has occurred; the new field line is within
\teq{2\rg} of the one the particle was circling before the  scattering.  Such
cross field diffusion is an integral part of diffusive acceleration at oblique
shocks (e.g. Jokipii 1987, Ellison, Baring, and Jones 1995), and its presence
is required in the Monte Carlo simulation in order to match  spacecraft
observations of particle spectra associated with  interplanetary shocks (Baring
et al. 1995).   Ellison, Baring and Jones (1995) showed that this scheme for
cross-field diffusion together with the assumption contained in
equation~(\ref{eq:mfp}) is equivalent to a kinetic theory description of
diffusion (e.g. Axford 1965; Forman, Jokipii and Owens 1974), where the
diffusion coefficients perpendicular to (\teq{\kappa_{\perp}}) and parallel to
(\teq{\kappa_{\parallel}}) the field are related via
\teq{\kappa_{\perp}=\kappa_{\parallel}/(1+\eta^2)}.  The parameter \teq{\eta}
in equation~(\ref{eq:mfp}) then clearly determines the strength of the
scattering  and when \teq{\eta \sim 1}, \teq{\kappa_\perp \sim
\kappa_\parallel}, the  so-called Bohm limit where particles diffuse across the
magnetic field as quickly as they move along it.  The properties of highly
oblique and quasi-parallel shocks tend to  merge when the scattering is strong.

\subsection{Grid Zone, Free Escape and Downstream Return Boundaries}

When a particle crosses a grid zone boundary, the values of the bulk flow
velocity and the magnetic field (both magnitude and direction for these vector
quantities) change and a new gyroradius, gyrocenter, and phase in the
gyro-orbit are determined, as outlined in the Appendix. The particle then
acquires a new gyromotion with subsequent convection along the new field
direction.   At a crossing of a grid zone boundary, we adopt the standard
requirement (e.g. see Terasawa 1979; Decker 1988; Begelman and Kirk 1990) in
orbit calculations that the momentum vector of the particle is conserved in the
de Hoffmann-Teller frame, since there is no electric field in the shock layer
in this frame.  While this implies conservation of energy in the HT frame (i.e.
between scatterings), particle energies do change at the discontinuity in the
NIF (e.g. Toptyghin 1980) due to the presence of drift electric fields, a
manifestation of the transformation between frames. This transmission criterion
differs from the  imposition of magnetic moment conservation that was made in
earlier applications of our Monte Carlo technique (e.g. Baring, Ellison and
Jones 1993); in  the present paper we make no  assumption concerning the
magnetic moment of a particle at a grid point or anywhere else. Particles may
of course `reflect' at any zone boundary or be transmitted depending on their
phase and pitch angle after a number of gyrations in the vicinity of the
boundary.

There are two limiting boundaries to the simulation region, the free escape
boundary (FEB) and the downstream return boundary. Our introduction of an
upstream FEB facilitates modelling of the finite extent, for example through
geometrical curvature, of  real shocks.  Escape is naturally expected in real
systems since the region of influence of a shock on its environs is finite, and
the level of shock-generated turbulence diminishes to background levels at
sufficient distances from the shock.   An escape boundary is most relevant to
the upstream region (i) because the direction of convection renders the
downstream region more spatially uniform, and (ii) because the upstream region
is usually on the convex side of the shock (e.g. supernova remnant shocks, the
Earth's bow shock), although particles can escape from the downstream side as
well.   The inclusion of such a free escape boundary is also motivated on
theoretical grounds.   The fundamental point is that for Fermi acceleration in
steady-state shocks, escape must occur for Mach numbers above some critical
value. This has been fully documented (e.g. Eichler 1984, 1985; Ellison and
Eichler 1984;  Jones and Ellison 1991), where it is observed that within the
context of the non-linear acceleration model, the Fermi
acceleration/hydrodynamics coupling becomes unstable for shocks of Mach numbers
above a few, and leads to singularities in the energy density when particle
escape is suppressed.  Finite solutions are achievable when an upstream FEB is
introduced, since its presence causes the acceleration process to truncate at
the highest energies.  In the case where the diffusion coefficient increases
with energy, the FEB produces a distribution which falls off approximately
exponentially at an energy where the upstream diffusion length is on the order
of the distance from the FEB to the shock.  In the simulation, particles are
removed just before they scatter for the first time on the upstream side of the
FEB; this choice leads to a spatial smearing of the effects of the FEB on the
scale length of the mean free path of the escaping particles, i.e. on length
scales comparable to the distance \teq{\dFEB} between the FEB and the shock. 
In the results presented here, the FEB is chosen close enough to the shock to
guarantee that all particles in the simulation remain non-relativistic; the
domain of acceleration to relativistic energies and also relativistic shock
scenarios are deferred to future work.  The dynamical consequences of the FEB
are discussed in detail in Section~4 below.

While the upstream region in the simulation is finite, delimited with a FEB, we
model an infinite downstream region with a probability of return calculation
beyond a downstream return boundary (DRB); this  spatial border renders the
simulation finite in time. Beyond the DRB, which is maintained more than a
scattering length downstream of the shock, the spatial diffusion properties are
treated using appropriate statistical probabilities.  If the position of a
particle (as opposed to its guiding center) is followed, then the probability
of return to the upstream side of the DRB assumes a simple form.  If the flow
is uniform with a component of velocity \teq{u_{\rm x2}} perpendicular to the
DRB (in our case, this direction is also perpendicular to the shock plane) and
particles of speed \teq{\vf} in the frame of the flowing plasma are also
isotropic in that fluid frame, then the probability, \teq{P_{\rm ret}}, that a
particle which crosses some arbitrary \teq{y}-\teq{z} plane will return to the
upstream side of that plane, is
\begin{equation}
   P_{\rm ret} = \left ( \dover{\vf -u_{\rm x2}}{\vf +u_{\rm x2}}\right )^2 \ .
  \label{eq:probret}
\end{equation}
While this calculation has been done many times (e.g. Bell 1978; Drury 1983;
Jones and Ellison 1991; and Ellison, Baring, and Jones 1995) we emphasize that
equation~(\ref{eq:probret}) is fully relativistic (Peacock 1981) and holds
regardless of the orientation of the magnetic field or the flow.  The principal
requirement for the validity of equation~(\ref{eq:probret}) is that the
particles are isotropic in the local fluid frame, a condition that is satisfied
since the DRB is at least a scattering length downstream of the shock.  Hence
while equation~(\ref{eq:probret}) can be used downstream where the flow is
uniform for any \teq{\vf\gtrsim u_{\rm x2}}, it cannot be used at the shock
where the flow speed changes unless \teq{u_{\rm x2}\ll\vf}.  The particle
speed, \teq{\vf}, must also remain constant during the time a particle spends
downstream from the \teq{y}-\teq{z} plane, a natural consequence of our elastic
scattering assumption.  The decision of return (or otherwise) is made via a
random number generator.  Particles which do return, must be injected back
across the \teq{y}-\teq{z} plane with properly flux-weighted $x$-components of
velocity, pitch angles, and phases.  The determination of these, along with a
detailed derivation of equation~(\ref{eq:probret}), are given in the Appendix. 
Note that the DRB is not only a feature of our Monte Carlo simulation, but is
also used in hybrid plasma simulations of shocks (Bennett and Ellison 1995).

\section{FLUX CONSERVATION RELATIONS AND SHOCK MODIFICATION}

Before presenting the results of our modified shock simulations, it is
instructive to review the elements of non-linear shock hydrodynamics
and our procedure for determining the fluid flow and magnetic field
spatial profiles that simultaneously conserve all relevant fluxes and
are also self-consistent products of the Fermi acceleration mechanism.

\subsection{Flux Conservation Relations}

The starting point for these considerations is the well-known one-dimensional,
steady-state, magnetohydrodynamic conservation relations (i.e., the
Rankine-Hugoniot (R-H) jump conditions) for an infinite, plane shock lying in
the \teq{y}-\teq{z} plane (see the geometry in Figure~\ref{fig:shock}). 
Variations of all quantities occur only in the \teq{x}-direction and these
equations are written in the normal-incidence-frame [NIF].  The notation is
that of Decker (1988), with the square brackets representing differences
between quantities far upstream (with the `1' subscript) and downstream (with
the `2' subscript) of the shock, however the origin of the forms used here is
based on the presentation on p.~56 of Boyd and Sanderson (1969).  For a
magnetic field strength of \teq{B}, if \teq{u} is the bulk speed of the plasma,
the purely electromagnetic equations (i.e. Maxwell's equations) are
\begin{equation}
   \left [ B_{\rm x} \right ]^2_1 \; =\; 0 \quad ,    \label{eq:divB}
\end{equation}
which defines a divergenceless magnetic field (remember that our system has 
\teq{\partial B_{\rm y}/\partial y= 0=\partial B_{\rm }z/\partial z}), and 
\begin{equation}
   \left [ u_{\rm z} B_{\rm x} - 
   u_{\rm x} B_{\rm z} \right ]^2_1 \; =\; 0 \quad ,      \label{eq:curlE}
\end{equation}
which expresses (since \teq{c\,\nabla \times {\bf E}=-\partial {\bf B}/
\partial t = {\bf 0}}) the uniformity of the tangential electric field across
the shock.  The hydrodynamic equations are as follows: the mass flux equation 
corresponding to the \teq{x}-direction is
\begin{equation}
   \bigl[ \rho u_{\rm x} \bigr]^2_1 \; =\; 0 \quad , \label{eq:massflux}
\end{equation}
where \teq{\rho} is the mass density; the equations for the flux in the
\teq{x}-direction of the \teq{x} and \teq{z} components of momentum are
\begin{equation}
   \biggl[ \rho u_{\rm x}^2 + P_{\rm xx} + { B_{\rm z}^2 \over 8\pi} 
   \biggr]^2_1 \; =\; 0 \quad ,                       \label{eq:xmoflux}
\end{equation}
and
\begin{equation}
   \biggl[ \rho u_{\rm x} u_{\rm z} + P_{\rm xz} 
   - \dover{B_{\rm x} B_{\rm z}}{4\pi} \biggr]^2_1 \; =\; 0\quad ,
  \label{eq:zmoflux}
\end{equation}
respectively, where \teq{P_{\rm xx}}  and \teq{P_{\rm xz}} are the appropriate
components of the pressure tensor.  Finally,  the energy flux in the
\teq{x}-direction satisfies
\begin{eqnarray}
   & & \Biggl[ \dover{\gamma}{\gamma - 1} P_{\rm xx} u_{\rm x} + 
   P_{\rm xz} \biggl\{ u_{\rm z} + \dover{u_{\rm x}}{3(\gamma -1)} 
   \biggl(\dover{2 B_{\rm x}}{B_{\rm z}} + \dover{B_{\rm z}}{B_{\rm x}}\biggr)
   \biggr\}\nonumber\\ 
   & & \qquad\qquad\qquad + \dover{1}{2}\rho u_{\rm x}^3 + 
   \dover{1}{2}\rho u_{\rm x} u_{\rm z}^2
   + { u_{\rm x} B_{\rm z}^2 \over 4\pi } - 
   \dover{u_{\rm z} B_{\rm x} B_{\rm z}}{4 \pi} 
   +Q_{\rm esc} \Biggr]^2_1 \; =\; 0 \quad .          \label{eq:enflux}
\end{eqnarray}
Here \teq{\gamma} is the ratio of specific heats, which enters via the thermal
contribution to the energy density using the equation of state; we set this
equal to 5/3 in this paper, since only non-relativistic particles appear in the
simulation results presented. Equations~(\ref{eq:xmoflux})--(\ref{eq:enflux}) 
neglect so-called gradient terms that are spatial diffusion contributions that
arise from non-uniformity of the fluid flow and magnetic field profiles.  Note
also that equations~(\ref{eq:xmoflux}), (\ref{eq:zmoflux}),
and~(\ref{eq:enflux}) approximate the respective parallel shock relations for
high Alfv\'enic Mach numbers (i.e.  where the field is dynamically
unimportant), whereas equation~(\ref{eq:curlE}) remains important regardless of
the Mach number.

In equation~(\ref{eq:enflux}) we have added the term,  \teq{Q_{\rm esc}}, to
model the  escape of particles at an upstream free escape boundary (FEB).  As
mentioned above, a FEB causes the acceleration process to truncate as particles
leave the system producing important dynamical effects since the escaping
energy, and therefore pressure, results in an increase in the compression ratio
of the shock (see Ellison, M\"obius, and Paschmann 1990 for a discussion of the
effects of such a term in the R-H relations).  The escaping energy flux,
\teq{Q_{\rm esc}}, is taken to be constant for the far downstream region and
zero for the region far upstream (i.e. several mean free paths) of the FEB, and
varies most rapidly in the neighbourhood of the FEB. We assume that the
concurrent escaping momentum and mass fluxes are small and neglect them in
equations~(\ref{eq:massflux}), ~(\ref{eq:xmoflux}), and~(\ref{eq:zmoflux}). 
This is a good approximation if the particles that escape have speeds such that
\teq{v_{\rm esc} \gg u_{\rm x1}} (see Ellison 1985), a situation that is always
realized in the simulation results presented here.

The appearance of different components of the pressure tensor in the
Rankine-Hugoniot relations is requisite for the Monte Carlo simulation since we
do not assume that particles are isotropic  in any frame.  Normally,
implementations of the conservation equations in astrophysical or heliospheric
applications (e.g. Decker 1988) are restricted to scenarios where the plasma is
isotropic in the local fluid frame, in which case \teq{P_{\rm xx}=P_{\rm
yy}=P_{\rm zz}=P} and the off-diagonal terms of the pressure tensor are zero. 
However, our system generates anisotropic plasma in all frames of reference,
due to the non-uniformity of the flow \it combined \rm with the
self-consistently determined Fermi acceleration of the particles.  In this
paper, for the sake of simplicity, we assume that the plasma is gyrotropic but
anisotropic in the local fluid frame in the flux equations.  

When an isotropic population of particles is convected across a velocity
discontinuity and then subjected to isotropic scattering, compression of the
plasma is close to, but not perfectly, gyrotropic: asymmetric phase sampling at
the discontinuity yields non-uniform phase distributions prior to scattering.
Plasma isotropy in the new local fluid frame is attained only after many
scattering lengths; in fact adjustment to true isotropy is never achieved in
our non-linear Monte Carlo treatment because the scale length of velocity (and
directional) changes in the fluid flow is always comparable to the
mean-free-path of the particles comprising the flow.  Gyrotropy is a good
approximation that is also expedient because it yields a diagonal pressure
tensor \teq{P_{\rm ij}} in the de Hoffman-Teller and fluid frames for
co-ordinate systems with one axis aligned along the magnetic field; since the
diagonal components generally differ, oblique shocks lead to non-zero
off-diagonal components: \teq{P_{\rm xz}= P_{\rm zx}\neq 0}.  A discussion of
the generation of such terms is presented in the Appendix, specifically
focusing on the details of the derivation of the pressure terms of the energy
flux in equation~(\ref{eq:enflux}); there the coefficient of \teq{P_{\rm xz}}
is alternatively expressed in terms of pressure components parallel to and
orthogonal to the local field.  Note that non-zero \teq{P_{\rm xy}=P_{\rm yx}}
also arise in hydrogenic plasma flows in conjunction with out-of-the-plane
components of the electric field (Jones and Ellison 1987, 1991); such
off-diagonal terms apply only to quantities in the $y$-direction, and therefore
are irrelevant to the considerations of this paper.

\subsection{Flux Scalings and Formalism}

The flux equations are used in the simulation in dimensionless form, scaling by
relevant upstream quantities.  In the NIF, the far upstream flow is taken along
the shock normal, i.e. \teq{u_{\rm z1} =0}.  We define the flow velocity
\teq{u_1\equiv u_{\rm 1x}}, the  magnitude of the far upstream magnetic field 
\teq{B_1\equiv\sqrt{B_{\rm 1x}^2+B_{\rm 1z}^2}\,},  and a far upstream plasma
density \teq{\rho_1}.  These specify a far upstream Alfv\'enic Mach number:
\teq{\MA=u_1/\vA} or \teq{\MA^2 = (4\pi \rho_1 u_1^2)/B_1^2}, where the Alfv\'en
speed is \teq{\vA=B_1/(4\pi \rho_1)^{1/2}}, and a far upstream sonic Mach
number, \teq{\MS^2 = \rho_1 u_1^2/(\gamma P_1)}, where \teq{P_1= n_1 \kB T_1}
is the far upstream (isotropic) pressure.  Here \teq{n_1} and \teq{T_1} are
the number density and temperature far upstream, and \teq{\kB} is Boltzmann's
constant.   These upstream parameters, along with \teq{\Tbn}, define the key
input for the simulation runs.  Using these definitions, one can write
equations~(\ref{eq:curlE}) and~(\ref{eq:xmoflux})--(\ref{eq:enflux}) in a
dimensionless form at any position \teq{x}:

\begin{equation}
   u_{\rm z}'(x) B_{\rm x1}^{'} - u_{\rm x}'(x) B_{\rm z}^{'}(x) \; =\; 
   \Fuxb \quad ,                                 \label{eq:curlEdim}
\end{equation}
defines the uniformity of tangential electric field, 
\begin{equation}
   u_{\rm x}'(x) + P_{\rm xx}^{'}(x)
 + { B_{\rm z}^{' 2}(x) \over 2 \MA^2} \; =\; \FpxxUpS \ ,
 \label{eq:xmodim}
\end{equation}
and
\begin{equation}
   u_{\rm z}'(x) + P_{\rm xz}^{'}(x)
 - { B_{\rm x1}^{'} B_{\rm z}^{'}(x) \over \MA^2} \; =\;\Fpxz \ ,
 \label{eq:zmodim}
\end{equation}
define the momentum flux equations, and
\begin{eqnarray}
  & & \dover{\gamma}{\gamma -1} P_{\rm xx}^{'}(x) u_{\rm x}'(x) + 
P_{\rm xz}^{'}(x) \left \{u_{\rm z}'(x) + 
  { u_{\rm x}'(x) \over   3 (\gamma -1)} 
  \left [ { 2B_{\rm x1}^{'} \over B_{\rm z}^{'}(x)} 
   + { B_{\rm z}^{'}(x) \over B_{\rm x1}^{'}} \right ] 
   \right \}\nonumber\\
& &\qquad\qquad + \dover{1}{2}u_{\rm x}^{' 2}(x) + \dover{1}{2}
u_{\rm z}^{' 2}(x) + {{B_{\rm z}^{'}(x)} \over {\MA^2}}
  \left[ u_{\rm x}'(x) B_{\rm z}^{'}(x) - u_{\rm z}'(x) B_{\rm x1}^{'} \right] 
  + Q_{\rm esc}^{'} \; =\; \FenUpS \ , \vphantom{\Biggl(} \label{eq:endim}
\end{eqnarray}
rearranges the energy flux equation.  All primed quantities are dimensionless,
using the notation \teq{u^\prime =  u/u_1}, \teq{B^{'} = B/B_1}, \teq{P^{'}=
P/(\rho_1 u_1^2)} and  \teq{Q_{\rm esc}^{'}=Q_{\rm esc}/(\rho_1u_1^3)}. Note
that \teq{\gamma=5/3} is a constant throughout the flow for the
non-relativistic applications here.  The constancy of the $x$-component of
magnetic field has been used to substitute  \teq{B_{\rm x}^{'}(x) =B_{\rm
x1}^{'}}, and we  have also used mass flux conservation
\begin{equation}
   \rho(x) u_{\rm x}(x) = \rho_1 u_1 \; =\; {\rm constant} \quad ,
\end{equation}
in these equations.  As mentioned above, the escaping momentum and mass fluxes
are of progressively smaller orders in \teq{u_{\rm x1}/v_{\rm esc}} than
\teq{Q_{\rm esc}^{'}}, and therefore are neglected.

The far upstream fluxes on the right hand sides of
equations~(\ref{eq:curlEdim})--(\ref{eq:endim})
are constants determined by the input shock parameters:
\begin{eqnarray}
   \Fuxb &=& - B_{\rm z1}^{'} \quad ,\nonumber\\
   \FpxxUpS &=& 1 + P'_1
     + { B_{\rm z1}^{' 2} \over 2 \MA^2} \quad ,\nonumber\\
   \Fpxz &=& - { B_{\rm x1}^{'} B_{\rm z1}^{'} \over \MA^2} 
                      \quad ,\nonumber\\
   \FenUpS &=& { \gamma \over \gamma - 1} P^{'}_1 + \dover{1}{2} +
      { B_{\rm z1}^{' 2} \over \MA^2} \quad .
\end{eqnarray}
Note that in all of the examples described below, we have set the
electron temperature equal to zero.  A finite electron temperature can
be modeled with our procedure and is necessary when fitting spacecraft
data (as done in Ellison, M\"obius, and Paschmann 1990), but is not
necessary for the discussion given here. We assume, as is generally
done, that the ions dominate the shock structure and the addition of
electrons has little effect other than changing the Mach number.

If \teq{Q_{\rm esc}^{'}} and \teq{P_{\rm xz}^{'}} are both assumed to be zero
at all \teq{x}, the four unknowns in
equations~(\ref{eq:curlEdim})--(\ref{eq:endim}); \teq{u_{\rm x}'(x)},
\teq{u_{\rm z}'(x)}, \teq{P_{\rm xx}^{'}(x)},  and \teq{B_{\rm z}^{'}(x)},  can
be obtained at every position \teq{x}.  This is just the standard situation of
a discontinuous shock, and these Rankine-Hugoniot relations are analytically
solvable (e.g. see Decker 1988) for the shock compression ratio, \teq{r\equiv
u_1/u_{\rm x2}}. However, in the modified collisionless shocks considered here,
the non-thermal component of the particle distribution that is generated by
particles crossing the shock more than once contributes significantly to the
total pressure of the system, and \teq{Q_{\rm esc}^{'}} will not, in general,
be zero. In fact, \teq{Q_{\rm esc}^{'}} will have different values at various
locations, and cannot be determined before the shock structure is known, making
a direct solution of equations~(\ref{eq:curlEdim})--(\ref{eq:endim})
impossible; this is the inherent non-linearity in the problem even if isotropy
is assumed,  defined by the coupling between the acceleration process and the
flow hydrodynamics.

In our approach, we iterate to achieve a solution for the velocity and field
shock profiles by varying \teq{u_{\rm x}'(x)},  \teq{u_{\rm z}'(x)},
\teq{B_{\rm z}^{'}(x)} {\it and} the  overall compression ratio, for successive
simulation runs (each accelerating particles and generating non-thermal
distributions) until equations~(\ref{eq:curlEdim})--(\ref{eq:zmodim}) are
satisfied at every \teq{x}.  The overall compression ratio depends on 
\teq{Q_{\rm esc}^{'}} far downstream  from the shock and is determined by our
solution.  When this value is consistent with equation~(\ref{eq:endim}), a
complete solution to the non-linear acceleration problem is obtained,
satisfying equations~(\ref{eq:divB})--(\ref{eq:enflux}) at all positions.  The
details of the iterative procedure follow.

\subsection{Iteration of the Shock Profile}

The iteration of the shock profile is done in two stages.  We first choose the
overall compression ratio (normally the R-H value for the first iteration) and,
using this ratio, we iterate the \it shape \rm of the profile.  As individual
particles move through the shock, the momentum and energy fluxes are calculated
at each grid zone boundary. We therefore obtain the quantity
\begin{equation}
\Fpxx = u_{\rm x}'(x) + P_{\rm xx}^{'}(x) + 
{ B_{\rm z}^{' 2}(x) \over 2 \MA^2}
\end{equation}
at each boundary, where \teq{u_{\rm x}'(x)} and \teq{B_{\rm z}^{'}(x)}  are the
current values for the shock structure. From this we compute the pressure
\teq{P_{\rm xx}^{'}(x)} and calculate a new \teq{x}-component of the flow speed
from equation~(\ref{eq:xmodim}),
\begin{equation}
u_{\rm x}^{'\rm N}(x) = \FpxxUpS - P_{\rm xx}^{'}(x) - 
 { B_{\rm z}^{' 2}(x) \over 2 \MA^2}
\end{equation}
such that the momentum flux will equal the constant far upstream value,
\teq{\FpxxUpS}, at all \teq{x}.  Replacing \teq{u_{\rm x}'(x)} with 
\teq{u_{\rm x}^{'\rm N}(x)}, we solve equations~(\ref{eq:curlEdim})
and~(\ref{eq:zmodim}) for new values of \teq{u_{\rm z}'(x)} and \teq{B_{\rm
z}^{'}(x)}.   To speed convergence, before running the next iteration we smooth
this new profile, force it to be monotonic, average it with the previous
profile, and scale the profile by setting all downstream values of the
\teq{x}-component of flow to \teq{u_{\rm x2}} and the far upstream value to
\teq{u_{\rm 1x}}.  Since no wave physics is  employed in our description, we do
not attempt to model anything other than a monotonic decrease in flow speed and
a monotonic increase in the magnitude and obliquity of \teq{{\bf B}} from
upstream to downstream.   Alternatively, we can calculate both \teq{P'_{\rm
xx}(x)} and \teq{P'_{\rm xz}(x)} in the simulation and obtain the new
prediction for \teq{u_{\rm x}^{'\rm N}(x)} from  equation~(\ref{eq:endim}), but
in either case, our procedure rapidly and stably converges.

With this new shock profile we repeat the simulation by again injecting
particles far upstream from the shock and propagating them until they leave at
either the FEB or the probability of return plane. Our algorithm converges
rapidly (within a few steps: see the examples in Section~4 below) to values of
\teq{u_{\rm x}'(x)}, \teq{u_{\rm z}'(x)},  and \teq{B_{\rm z}^{'}(x)}, which no
longer change significantly with subsequent iterations. However, in general,
this profile will not simultaneously conserve momentum and energy fluxes unless
a compression ratio consistent with the escaping energy flux \teq{Q_{\rm esc}}
has been chosen. The second stage of the iteration process is to successively
choose new overall compression ratios (larger than the R-H value), each time
repeating the iteration of the profile shape, and continue until the momentum
fluxes and the energy flux (with \teq{Q_{\rm esc}^{'}} added) are constant
everywhere.  Thus, within statistical limits, a shock profile and overall
compression ratio that are consistent with the Fermi acceleration process are
obtained.

\section{RESULTS}

Oblique shocks are highly complex, even in the steady state and in plane
geometry, and several parameters control the dissipative processes as well as
the injection from thermal energies into the Fermi acceleration mechanism. 
These far upstream parameters include
the magnetic field strength, \teq{B_1},
the obliquity, \teq{\Tbnone},
the temperature, \teq{T_1},
the number density, \teq{n_1},
and the shock speed, \teq{u_1},
all of which are determined by the ambient upstream conditions and can, in
principle, be determined by observations of a given physical system.  The size
of the acceleration region is also an observable (for example the radius of a
supernova remnant shock), and we model it using the distance \teq{\dFEB}
between the upstream free escape boundary and the shock. However, the `size' of
the shock in units of mean free paths is very important and this will depend on
the scattering law we assume.  This requires the introduction of another
parameter, \teq{\eta}, the ratio of the mean free path to the gyroradius, via
equation~(\ref{eq:mfp}). The value of \teq{\eta}, which determines the amount
of cross-field diffusion, depends on the highly complex plasma interactions
that occur in the shock environs; the prescription in equation~(\ref{eq:mfp})
is a simple but insightful way to model these plasma processes.

Another ``variable'' results from the inclusion of two extreme modes of
scattering, namely large-angle scattering (LAS) and pitch-angle diffusion
(PAD).  While more complicated scattering models can be used, we believe these
contain the essential physics of plane shocks and yield important information
on the nonlinear processes linking shock structure and particle acceleration. 
The type of scattering we employ and \teq{\eta} are free parameters and cannot
be determined in our model except by comparison with observations of space
plasma shocks or 3-D plasma simulations.  Hence (replacing \teq{B_1} and
\teq{T_1} with \teq{\MA} and \teq{\MS}), there are seven parameters in our
model: \teq{\MA}, \teq{\MS}, \teq{n_1}, \teq{\Tbnone}, \teq{u_1}, \teq{\dFEB},
and \teq{\eta}, together with the choice of the type of scattering (either LAS
or PAD).  In all of the following examples we use a shock speed of \teq{u_1=500
}km~sec$^{-1}$ and a far upstream number density of \teq{n_1 = 1}cm$^{-3}$,
which define physical scales for our system that are more or less appropriate
for astrophysical shocks. The spatial scales of the results presented here are
all in units of the ``low-energy mean free path'' \teq{\lambda_0}, which is
defined as the mean free path [see  equation~(\ref{eq:mfp})] of a proton with
speed \teq{u_1} in the upstream magnetic field, i.e., \teq{\lambda_0=\eta
m_{\rm p} u_1 c/(e B_1)}, where \teq{\eta} (\teq{\geq 1}) remains an adjustable
parameter for each simulation run.

\subsection{Test-Particle Examples}

The simplest acceleration results obtainable from the Monte Carlo simulation
are for test-particle cases where the shock profile is uniform on either side
of the subshock; this limit corresponds to the first run in the iteration
sequence described in Sections~3.2 and~3.3, and has been studied in detail in
Ellison, Baring and Jones (1995). Several interacting elements of the code,
including shock (or grid zone) crossings and the probability of return
calculation (which includes the flux-weighting of momenta of the returning
particles: see Section~A.3 in the Appendix) must be implemented properly in
order to yield the well-known test-particle acceleration power-law.  The Fermi
power-law is achieved when particle speeds \teq{v} far exceed the HT flow speed
\teq{u_{\rm x1}/\cos\Tbnone }.  From analytic calculations 
(e.g. Drury 1983; Jones
and Ellison 1991), the spectral index \teq{\sigma} of the power-law then
depends only on the compression ratio \teq{r=u_1/u_{\rm 2x}} regardless of the
obliquity or other plasma parameters. The compression ratio is determined from 
equations~(\ref{eq:divB})--(\ref{eq:enflux}) with \teq{Q_{\rm esc} = P_{\rm xz}
=0} (see also Decker 1988). For non-relativistic particle energies and shock
speeds the  test-particle distribution is:
\begin{equation}
   \dover{dJ}{dE} \propto E^{-\sigma}\quad ,\quad \sigma\; =\;
   { r+2 \over 2(r-1)} \quad ,                   \label{eq:testpower}
\end{equation}
where \teq{dJ/dE} is the number of particles in units of 
(cm$^{2}$-sec-steradian-keV)$^{-1}$,  i.e. is an omni-directional flux (see
Jones and Ellison 1991).  The reproducibility of this form is a powerful tool
for debugging the portions of the code that are directly related to the
transport and acceleration of particles.  It is instructive to review
test-particle distributions before proceeding to our results for the non-linear
problem.

\placefigure{fig:testspectra}          

In Figure~\ref{fig:testspectra} we show spectra calculated with a discontinuous
shock for three different sets of parameters and for both large-angle
scattering (solid lines) and pitch-angle diffusion (dotted lines).  Note that
the examples labelled (c) are multiplied by 0.01 for clarity of display.  All
spectra here and elsewhere are omni-directional, calculated several
\teq{\lambda_0} (i.e. ``thermal'' mean free paths) downstream from the shock in
the normal incidence frame, and are normalized to one particle per square cm
per second injected far upstream.  In all cases we have used
\teq{u_1=500}km~sec$^{-1}$, and \teq{n_1=1}cm$^{-3}$; the (sonic and
Alfv\'enic) Mach number 20 cases here use \teq{B_1= 1.15\times 10^{-5}} G and 
\teq{T_1=4.54\times 10^{4}} K, while the Mach number 3 case uses  \teq{B_1=
7.64\times 10^{-5}} G and \teq{T_1=2.02\times 10^{4}} K. The other parameters
are given in the figure.  No free escape boundary is included in these
test-particle cases, being necessary only for the non-linear acceleration
problem.  

Note that the compression ratio varies slightly between the two high Mach
number examples, being \teq{r=3.96} for the \teq{\Tbnone=30^\circ} case and
\teq{r=3.94} for the \teq{60^\circ} case; this occurs because of a slight
dependence of \teq{r} on obliquity for non-infinite Mach numbers.  For the low
Mach number example (c), the shock is quite weak with \teq{r=2.5}.  The Fermi
power-laws obtained from equation~(\ref{eq:testpower}) for these compression
ratios are shown as light solid lines in the figure (with adjusted
normalization to aid visual distinction).  Clearly the most important feature of
Figure~\ref{fig:testspectra} is that the simulation does reproduce the Fermi
power-law index at high energies for a wide range of shock parameters.
Comparison of examples (a) and (b) with similar compression ratios but quite
different \teq{\Tbnone}, supports the fact that the Fermi spectral index
\teq{\sigma} is determined solely by \teq{r}.

The next most striking feature of these plots is that the two modes of
scattering produce very little difference in the spectra.  This difference is
largest for portions of the \teq{\Tbnone=60^\circ} spectrum (b) between thermal
energies and about \teq{100}keV. There the distribution for the LAS mode [the
solid line in (b)] is somewhat noisy due to comparatively poor statistics in
high \teq{\eta}, high obliquity runs (i.e. weak injection: see Ellison, Baring
and Jones 1995), and this noise may obscure some underlying structure that can
arise from large energy boosts in individual shock crossings.  At these
energies, the particle speed does not far exceed the HT frame flow speed
\teq{u_{\rm x1}/\cos\Tbnone}, 
leading to  significant anisotropies in the particle
population, and more importantly, measurable differences in the degree of
anisotropy produced in PAD and LAS modes.  Since such differences in angular
distributions are responsible for observed differences between LAS and PAD
applications to unmodified relativistic shocks (e.g. see Ellison, Jones, and
Reynolds 1990 for a comparison of the modes, and Kirk and Schneider 1987;
Ostrowski 1991 for PAD cases), it is not surprising that spectral differences
should appear here at suprathermal energies for highly oblique shocks. 
Clearly, Figure~\ref{fig:testspectra} shows that for this intermediate
obliquity case, the mode of scattering has virtually no effect on the
resultant spectrum at either thermal or the highest energies, and therefore
that the scattering mode plays little role here in determining the
efficiency of acceleration, i.e. the fractional energy deposited in high energy
particles.  This is not surprising, because of the low value of \teq{\eta}
here: we naturally expect that strong scattering (near the Bohm diffusion
limit) will destroy any sensitivity of the acceleration process to the shock
obliquity or distribution anisotropies, and hence the type of scattering.
Later, we shall see that larger (but still relatively small) differences between
the scattering modes occur for weak scattering (i.e., large \teq{\eta}). 

Also evident in Figure~\ref{fig:testspectra} is the strong effect the input
parameters have on injection efficiency: the \teq{\Tbn=60^\circ} spectra fall
an order of magnitude below the \teq{\Tbn=30^\circ} spectra at high energies
even though they both obtain the same power law index.  Increasing either
\teq{\Tbnone} or \teq{\eta} will result in decreased injection efficiency. 
These effects were detailed in Ellison, Baring, and Jones (1995), where an
anti-correlation between acceleration time and efficiency of acceleration in
test-particle shocks was observed.  We note that existing analytic predictions
for the transition between the thermal peak and the high energy power-law,
i.e., the injection efficiency, require {\it ad hoc} parameters additional to
and independent of those made for the shock structure to connect the thermal
gas to the cosmic ray population (e.g. Zank, Webb, and Donohue 1993; Kang and
Jones 1995; Malkov and V\"olk 1995).  The advantage of our model is that the
single relation [equation~(\ref{eq:mfp})] controls the shock structure, the
absolute injection efficiency, and, in fact, the entire shock solution.  This
makes it straightforward to compare model predictions to observations and to
infer plasma properties (such as the level of turbulence, the correctness of
the elastic scattering assumption, etc.) from these comparisons, an attractive
feature.  Properties of upstream particle distributions are deferred to the
discussion of non-linear results in the next subsection, though test-particle
spectra upstream of oblique shocks were presented in Baring, Ellison and Jones
(1994).

\subsection{Examples Showing Iteration of Shock Profile}

For our next examples, we compute the self-consistent smooth shock profile
beginning with a low Mach number case, i.e., \teq{\MA=\MS=3}, 
\teq{\Tbnone=30^\circ}, \teq{\dFEB=-50\lambda_0}, and \teq{\eta=2}, yielding a
plasma \teq{\beta} of \teq{\beta_1=1.2}.  This corresponds to a weak shock,
typical of interplanetary shocks observed in the heliosphere (e.g. see Burton
et al. 1992, Baring et al. 1995).  To reiterate, in the results that follow,
all lengths are measured in units of \teq{\lambda_0}, which is the mean free
path  \teq{\eta r_{\rm g1}} of a proton of gyroradius \teq{r_{\rm g1}=m_{\rm p}
u_1 c/(e B_1)}, i.e. with the speed \teq{u_1} in the upstream magnetic field.

In the left panels of Figure~\ref{fig:twoprofiles} we  depict the average flow
speed, \teq{u_{\rm x}}, the flux \teq{\Fpxx} of the $x$-component of momentum,
and the energy flux \teq{\Fen}, all normalized to far upstream values,  for
several iterations  starting from a discontinuous shock (light solid line) and
yielding the final profile (heavy solid line).  These iterations were done
using LAS and an overall compression ratio of \teq{r \simeq 2.7} which was
determined in previous iterations on \teq{r}.  The Rankine Hugoniot compression
ratio with \teq{Q_{\rm esc}=0} is \teq{r=2.67} which is equal (within errors)
of our 2.7 value.  The convergence is quite rapid and the heavy solid lines
(fourth iteration) show no further statistically significant change with
additional iterations.  Except for a departure of about 5\% near \teq{x=0}, the
flux of the $x$-component of momentum (middle panels) is constant for all
\teq{x} after the final shock profile has been obtained.  The flux \teq{\Fpxz}
of the $z$-component of momentum is generally small and less interesting; its
profile (and those of \teq{u_{\rm z}} and \teq{B_{\rm z}}) is not displayed for
reasons of brevity.  For the discontinuous shock (light solid lines), the
momentum flux was clearly not conserved and rose to \teq{\sim 140\%} of the far
upstream value downstream from the shock.

\placefigure{fig:twoprofiles}          

The escaping energy flux at the FEB (which is at $-50\lambda_0$ and not shown
in the figure), is less than 1\% of the far upstream value and does not
influence the overall compression ratio significantly.  The check on the
consistency of the final profile is that the momentum flux and the energy flux,
including the escaping flux, must both be conserved.  When this is achieved
(i.e.  corresponding to the solid lines), we have a unique, self-consistent
solution.  The \teq{\sim 10\%}  discrepancy in the energy flux  near \teq{x=0}
is most likely the result of the strong gradients in the shock and/or 
agyrotropic pressure tensor terms in the Rankine-Hugoniot relations which we
have neglected from our flux considerations.  This discrepancy decreases
rapidly with increasing Mach number, and so is greatest for this present case.

Figure~\ref{fig:twoprofiles} also shows the shock structure obtained with
pitch-angle diffusion PAD (right panels), all other shock parameters being the
same as the LAS case.  Within statistics, the two scattering modes give
identical results, both for the shape of the profile and the overall
compression ratio.  This feature is not surprising given that the test-particle
results of the previous simulation bore this similarity out.  

\placefigure{fig:specone}          

In Figure~\ref{fig:specone} we show the distribution functions generated by the 
smooth shocks of Figure~\ref{fig:twoprofiles}.  The spectra are
omni-directional, measured downstream from the shock, and calculated in the
shock (i.e. NIF) frame.  As with the shock structure, the distribution functions
obtained with the two scattering modes are identical within statistics.  This is
to be expected because \teq{\eta =2} is close to the Bohm diffusion limit where,
as mentioned above, isotropic diffusion will naturally obscure the differences
between PAD and LAS. There is a large difference, however, between the smooth
shock results and the test-particle, discontinuous shock (dotted line), which
was obtained using the same compression ratio (\teq{r=2.7}) computed in the
self-consistent solution of the non-linear LAS simulation. The light solid line
is the Fermi power-law expected from \teq{r=2.7}; the test-particle spectrum
attains this result before the falloff at \teq{\sim 100}keV produced by the FEB.
The discontinuous shock produces more efficient acceleration at low energies
than the smooth shock, which follows from the nature of the smoothed shock
profile: low energy particles feel the effect of the subshock whereas only the
high energy particles sample the full compression ratio of \teq{r=2.7}.  This
same property is responsible for the upward curvature of the non-linear spectra,
which never attain the Fermi power-law, but flatten towards it before falling
off due to the FEB.

We have also obtained solutions (not shown) using exactly the same parameters as
above except with a FEB at \teq{\dFEB=-10\lambda_0}. The smaller shock system
causes a cutoff at a lower energy than seen in  Figure~\ref{fig:specone} and the
shock structure is correspondingly on smaller length scales.  However, the
self-consistent compression ratio is still \teq{r\sim 2.7}, consistent with a
\teq{Q_{\rm esc}=0}, as expected, since these low Mach number shocks put a small
fraction of the available energy into energetic particles regardless of the
shock size. 

\placefigure{fig:febprofile}          

As a more extreme example, we show in Figure~\ref{fig:febprofile} a high Mach
number shock (\teq{\MS=\MA=20}) with a much larger shock size, i.e., the FEB is
at \teq{\dFEB=-200\lambda_0}.  The other input parameters are:
\teq{u_1=500}km~sec$^{-1}$, \teq{n_1=1}cm$^{-3}$,  \teq{\Tbnone=30^\circ},
\teq{\eta=2},  \teq{B_1=1.15\times 10^{-5}} G, and  \teq{T_1=4.54\times 10^{4}}
K, yielding  \teq{\beta=1.2}.  We have only used the LAS scattering mode since
the PAD results are essentially identical.   Note that the self-consistent
compression ratio used here is \teq{r=5} (well above the R-H value of
\teq{r=3.96})  which has been determined with previous runs not shown. The
distance scale in Figure~\ref{fig:febprofile} is logarithmic for
\teq{x <-10\lambda_0} and linear for \teq{x > -10\lambda_0}.

There are several important features of this shock solution.  In the first
iteration with no shock smoothing, the momentum and energy fluxes are wildly
non-conserved with both of them obtaining downstream fluxes almost 15 times as
large as the upstream values.  Despite this, the subsequent iterations converge
rapidly and by the fourth iteration the momentum flux is conserved everywhere
to within 5\% of the upstream value (the first, second, third, and fourth
iterations are shown by light solid, dashed, dash-dot, and heavy solid lines,
respectively; the flat dotted line indicates the far upstream value). The
effects from the anisotropic terms in the momentum and energy fluxes are less
noticeable here than in the previous low Mach number examples.  As with the
previous examples, the shock is smoothed out to the FEB, however the subshock
here is considerably more distinct showing a sharp discontinuity between the
flow just upstream from the subshock and the downstream flow.  The width of the
subshock is well within one \teq{\lambda_0}.

As for the energy flux, it falls about 20\% below the far upstream value due to
the particles lost at the FEB.  The escaping energy flux, \teq{Q_{\rm esc}},
which is zero far upstream, falls rapidly around  \teq{\dFEB} and then becomes
approximately constant into the downstream region. This results in a
compression ratio, obtained by iteration in previous runs,  of \teq{r\simeq 5},
compared to the R-H value of \teq{r=3.96}. A compression ratio of \teq{r=5}
implies  \teq{Q'_{\rm esc}=0.084}  and a \teq{Q'_{\rm esc}/F'_{\rm en1}\simeq
0.17},  and when this is added to the energy flux shown in  the bottom panel of
Figure~\ref{fig:febprofile}, we have a self-consistent solution with all fluxes
constant at all \teq{x} to within a few percent. Larger escaping fluxes are
expected for such high Mach number shocks because their greater compression
ratios enhance the acceleration efficiency to the highest energies.

\placefigure{fig:febstruct}          

The complete shock structure is shown  in Figure~\ref{fig:febstruct},  where we
have included along with \teq{u_{\rm x}(x)}, the \teq{z}-component of flow
speed, \teq{u_{\rm z}(x)},  the angle the local magnetic field makes with the
shock normal, \teq{\Theta_{\rm Bn}(x)}, and the total magnetic field magnitude,
\teq{B_{\rm tot}(x)}. The solid lines show the final shock structure and this is
compared to the initial structure shown by dotted lines.  As noted above, both
the shape and the overall compression ratio must be modified to obtain a
self-consistent solution, and this translates to a change from the R-H values
in the downstream \teq{u_{\rm z2}}, \teq{\Tbntwo}, and \teq{B_2} as indicated
by the dotted lines.  The sharpness of the subshock in
Figure~\ref{fig:febstruct} at \teq{x=0} is partially an artifact of how we
smooth and truncate the flow profile between iterations. As mentioned above, we
average the predicted profile with the previous one and, since we start with a
discontinuous shock, some sharpness persists.  We also set all predicted values
of \teq{u_{\rm x}} to \teq{u_{\rm x2}} at \teq{x>0}.  Despite this, the
subshock is, in fact, quite sharp as we discuss at the end of this section.

\placefigure{fig:febspec}          

In Figure~\ref{fig:febspec} we show (solid line) the downstream, shock frame
distribution function obtained in the smooth shock solution just described.  It
differs considerably from the test-particle solution (dotted line) in that the
downstream thermal peak is at a lower energy, the  temperature is slightly
lower, and far fewer thermal particles become accelerated. The straight line is
the power-law slope expected from the test-particle Fermi solution with
\teq{r=5}  and matches our test-particle solution at energies well above
thermal and below where the FEB becomes important.  The smooth shock solution,
however, does not attain the test-particle power-law and remains considerably
steeper.  This can be understood by examining the top panel of
Figure~\ref{fig:febprofile} where it can be seen that  at \teq{\dFEB
=-200\lambda_0} there is enough escaping energy flux  to smooth the shock
further upstream from this point.  Particles leave the shock at the FEB before
feeling the full compression ratio.

\placefigure{fig:upsspec}          

To complete this example, we show  in Figure~\ref{fig:upsspec} distribution
functions at various \teq{x}-positions, i.e.,    \teq{x= -50 \lambda_0} (dotted
line), \teq{x= -4 \lambda_0} (dashed line), \teq{x=-0.5 \lambda_0} (light solid
line), and  \teq{x=+\lambda_0} (heavy solid line).  At observation points far
upstream from the shock, only the unshocked thermal peak is present since
energetic particles from the shock are not able to diffuse against the
background flow to reach the observation point.  As the upstream observation
point moves toward the shock two things happen.  First, the highest energy
particles begin to show their presence, and second, the thermal peak from
particles which have not yet crossed the shock begins to shift to lower energy
(see insert where we have plotted the thermal peaks of the  \teq{x= -50
\lambda_0} and \teq{x=-0.5 \lambda_0} spectra).  Since we take \teq{\lambda
\propto \rg}, the diffusion length increases with energy and higher energy
particles from the shock are able to stream further upstream than low energy
ones, thus as the observation point moves toward the shock the spectrum fills
in from high energy to low (this property was recognized, for test-particle
situations, by Baring, Ellison and Jones 1994).  The shift in the  thermal peak
arises because  spectra calculated in the shock frame shift to lower
energy as the bulk flow speed falls as the shock is approached.  The slowing of
the bulk flow in the shock precursor also heats the incoming particles somewhat
before they encounter the sharp subshock lowering the local Mach number.  These
features are well-defined model predictions that can be tested against
observations.

\placefigure{fig:upsscale}          

In Figure~\ref{fig:upsscale} we show the upstream scale height for particles of
various energies.  The ordinate is the ratio of the flux at \teq{x} over the
flux at \teq{x=0} for a given energy.  As expected, the length scale  (i.e.
distance at which the flux \it e-folds\rm ) is largest for the highest energy
particles, and the fluxes fall off exponentially with distance from the shock (a
property of diffusion against the convecting flow). It is also important to note
that low energy particles (i.e. the 3 and 10 keV examples) can have extremely
short upstream precursors.  Particle detectors on spacecraft being overtaken by
interplanetary shocks will see very different time profiles depending on the
particle energy sampled and may, depending on the time integration of the
spectrometer (which is usually long compared with typical gyroperiods), see a
step function increase in intensity at low energies simultaneously with a slow
rise in high energy particles.  This effect may explain some puzzling aspects of
recent Ulysses observations which have led to the suggestion that a two-stage
acceleration mechanism operates for pickup protons (Gloeckler et al. 1994). 

\placefigure{fig:sfprofile}          

As our final example, we show  in Figure~\ref{fig:sfprofile} a highly oblique
shock (\teq{\Tbnone =75^\circ}) with \teq{\MS =\MA =10}, \teq{\dFEB=-20
\lambda_0}, \teq{\eta=5}, and using large-angle scattering.  For these
parameters little acceleration occurs and the shock profile is nearly
discontinuous.  Nevertheless, the little smoothing evident in the top panel of
Figure~\ref{fig:sfprofile}  is enough to reduce the energy flux from being
\teq{\sim 20\%} above the far upstream value to a constant value.  Because of
the inefficient acceleration, few particles, carrying very little energy flux,
escape at the FEB, so there is no need to adjust the compression ratio. The
final ratio is \teq{r=3.74}, effectively the R-H value. We have done the same
calculation with the same parameters except using pitch-angle diffusion and
find essentially the same profile (which is not shown), although some slight
differences do show up in the distribution functions.

\placefigure{fig:sfspec}          

The top two curves in Figure~\ref{fig:sfspec} are the distribution function
from the LAS example (solid line) along with the distribution produced in a
shock with the same parameters as that shown in Figure~\ref{fig:sfprofile},
only using PAD (dashed line). The main difference between the two cases is that
the PAD distribution is somewhat smoother than the LAS one. The PAD shock is
also somewhat less efficient in accelerating particles to the highest energies. 
In general, however, even at this large obliquity, the choice of scattering
mode does not play a dominant role in determining the acceleration efficiency.
The lower two curves (multiplied by 0.01 for clarity) are test-particle results
and are similar to the distribution functions produced by the smooth shock, but
slightly flatter, as expected.  No portions of these examples elicit the
Fermi, test-particle power-law slope (dotted line) but are clearly flattening
toward it before the FEB causes the spectra to turn over.

\placefigure{fig:aveflow}          

Finally, we comment on the sharpness of the subshock seen in all of the
examples we have presented.  As mentioned above, part of this is due to the
scheme we have for iterating the profile and insuring rapid convergence.  We
average the predicted profile with the previous one, make the predicted profile
monotonic, and set all predicted values of \teq{u_{\rm x}(x)} to \teq{u_{\rm
x2}} for \teq{x>0}.   In Figure~\ref{fig:aveflow} we show the same final
\teq{u_{\rm x}(x)}  plots as shown in the top left panel of
Figure~\ref{fig:twoprofiles}, the top panel of Figure~\ref{fig:febprofile}, and
the top panel of Figure~\ref{fig:sfprofile}, except here they are all plotted
on an expanded linear distance scale (solid lines). The dotted lines in
Figure~\ref{fig:aveflow} are the \it calculated \rm \teq{<\!\! v_{\rm x}\!\!>}
obtained from distributions produced by the simulation using the solid line
shock profiles. In all cases, the mean velocities are nearly as sharp as our
processed profiles except for statistical fluctuations and the truncation we
impose at \teq{x > 0}.  Outside of the range shown, \teq{u_{\rm x}} and
\teq{<\!\!v_{\rm x}\!\!>} are indistinguishable except for noise. With the
possible  exception of the low Mach number case, a clear subshock exists which
is considerably less than \teq{\lambda_0} in width.   While comparing
\teq{u_{\rm x}} and \teq{<\!\! v_{\rm x}\!\!>} is somewhat artificial since
\teq{<\!\! v_{\rm x}\!\!>} is calculated using \teq{u_{\rm x}}, the small
broadening of the flow speed around \teq{x\sim 0} doesn't have an appreciable
influence on the momentum and energy fluxes. We have performed simulations
using \teq{<\!\! v_{\rm x}\!\!>} instead of  \teq{u_{\rm x}(x)} and the
momentum and energy fluxes remain within \teq{\sim 10\%} of the conserved
values at all \teq{x}. For our purposes, since we do not attempt to model
scales less than a convected thermal ion gyroradius, there is essentially no
difference between the \teq{u_{\rm x}(x)}  used in the simulation and
\teq{<\!\! v_{\rm x}\!\!>}. 

\subsection{Injection and Acceleration Efficiency}

A quantity that is central to the acceleration problem is the efficiency of the
Fermi mechanism.  It can be defined in a variety of ways: we define the
acceleration efficiency, \teq{\epsilon (>\! E)}, at or behind the shock as the
downstream energy flux above energy \teq{E} divided by the incoming energy
flux, i.e.,  
\begin{equation}
  \epsilon (>\! E) \; =\; {\zeta P(>\! E) u_{\rm x2} + Q_{\rm esc}
  \over \zeta P_1 u_1 + \rho_1 u_1^3/2 + B_{\rm z1}^2 u_1/(4\pi)}
  \quad , \quad \zeta\; =\; \dover{\gamma}{\gamma -1}\quad ,
 \label{eq:acceff}
\end{equation}
where \teq{P(>\! E)} is the downstream pressure in particles with energies
greater than \teq{E}.  The pressure is obtained by
taking \teq{2/3} of the energy density in the omni-directional distribution,
\teq{dJ/dE}, and we take \teq{Q_{\rm esc}} to be energy independent, i.e.,  we
assume all of the escaping energy flux is carried by the highest energy
particles.  This last assumption will distort \teq{\epsilon (>\! E)} somewhat
at the highest energies since particles of varying energies leave the FEB.

\placefigure{fig:figacceff}          

In Figure~\ref{fig:figacceff} we show \teq{\epsilon (>\! E)} for our three
previous non-linear examples, i.e.,  curve (a) is for the LAS case shown in
Figure~\ref{fig:specone},  (b) is for the case shown in
Figure~\ref{fig:febspec}, and (c) is for the LAS, smooth shock case shown in
Figure~\ref{fig:sfspec}.  As is apparent from the figure, large differences in
the efficiency depending on \teq{\Tbnone}, the Mach number, \teq{\eta}, and the 
distance to the FEB occur. Examples (b) and (c) both accelerate particles to
above 1000 keV and have similar Mach numbers, but the highly oblique shock (c),
is much less efficient.  Examples (a) and (b) have the same
\teq{\Tbnone=30^\circ} but differ considerably in Mach numbers (3 versus 20,
respectively), with (b) being more efficient because its higher Mach numbers
generate a larger compression ratio.  At this stage of our work, there appears
to be no simple way to characterize the injection and acceleration efficiency
of oblique shocks, but the trends with obliquity and Mach number are quite
clear. The flattening at the highest energies seen in (a) and (b) is an
artifact of assuming that \teq{Q_{\rm esc}} is carried totally by the highest
energy particles; if a spread of escape energies were possible, corresponding
to a range of \teq{\dFEB}, the flattening would be replaced by a
quasi-exponential turnover.  Note also, that the efficiency curves in 
Figure~\ref{fig:figacceff} do not quite converge to unity at low energies due
to simulation statistics.

Another point which should be made concerning Figure~\ref{fig:figacceff} is that
we choose to define our efficiency as a function of energy. If we do not have an
independent source of energetic seed particles, all accelerated particles must
originate as thermal particles and they will be drawn more or less continuously
from the thermal population -- there will be no clear separation between thermal
and energetic particles.  Our identical treatment of the thermal and non-thermal
populations is why we have defined our efficiency as a function of energy. 
However, there are ways to qualitatively prescribe efficiencies that describe
the overall distribution rather than different particle energies.  For example,
while all three cases in Figure~\ref{fig:figacceff} have comparable efficiency
at around 0.8 keV, this energy does not define the  approximate juncture between
thermal and non-thermal populations for cases (a) and (c), whereas it does for
case (b).  This juncture is  \it roughly \rm represented by the upward kinks at
energies 3 keV and 1.7 keV for cases (a) and (c) respectively.  Hence, the ratio
of downstream non-thermal to thermal energy densities for the three cases are
roughly (a) 0.2, (b) 0.7 and (c) 0.2; these numbers could be taken as an
alternative measure of acceleration efficiency.

\section{DISCUSSION AND CONCLUSIONS}

A host of observational evidence, both direct and indirect, confirms that
collisionless shocks in space accelerate particles with high efficiency.
Possibly a large fraction of {\it all} non-thermal particle populations in
diffuse regions of space are generated by shocks, making shock acceleration one
of the most important problems in high energy astrophysics.  As a step toward a
full understanding of shock acceleration, we have developed a model which
combines nonlinear particle acceleration  and diffusion with shock dissipation
and non-linear hydrodynamics forming the shock structure. This paper presents
both the details of our simulation technique and   representative acceleration
results as a prelude to a more comprehensive survey of the parameter space
associated with modified, oblique shocks. While our model is still incomplete,
with simplifying assumptions concerning the microphysical processes involved,
we believe it is the most realistic current solution of the steady-state shock
acceleration problem.  We include (i) a strongly energy-dependent diffusion
coefficient [see equation~(\ref{eq:mfp})], which models cross-field diffusion, 
(ii) the ability to model either large-angle scattering or pitch-angle
diffusion, (iii) injection from the thermal background with no additional free
parameters,  (iv) the determination of the self-consistent, average shock
structure including the dynamic effects of accelerated particles on the thermal
shock, (v) the dynamic effects of particle escape from finite shocks, and 
(vi) shock drift and compressional acceleration simultaneously.  Principal
results of this paper include downstream spectra (see Figures~\ref{fig:specone},
\ref{fig:febspec} and~\ref{fig:sfspec}), properties of upstream populations
(see Figures~\ref{fig:upsspec} and~\ref{fig:upsscale}), and acceleration
efficiencies (Figure~\ref{fig:figacceff}). The Monte Carlo simulation does not
treat (i) a self-consistent determination of the  diffusion coefficient from
wave-particle interactions, (ii) time-dependent effects, (iii) relativistic
particles or flow speeds (iv) a cross-shock potential due to charge separation,
or (v) geometry other than a plane shock.

Clearly the most important omission is the self-consistent determination of the
diffusion coefficient.  Our results hinge on the energy-dependent form we have
assumed for \teq{\kappa}, which is motivated by previous theoretical analyses
(see Jokipii 1987; Giacalone, Burgess and Schwartz 1992; Ellison Baring and
Jones 1995), and also observational constraints at parallel shocks (see
Ellison, M\"obius, and Paschmann 1990 and Ellison and Reynolds 1991).  However,
the self-consistent determination of \teq{\kappa} requires knowledge of the
microphysics and only plasma simulations where the electric and magnetic fields
are calculated directly from particle motions (e.g. Quest 1988) can give this
information.  These simulations are extremely demanding computationally and will
not, in the foreseeable future, be able to adequately model particle
acceleration in shocks to astrophysically important energies. The recent work
of Jokipii et al. (Jokipii, Giacalone, and K\'ota 1993; Giacalone, Jokipii, 
and K\'ota 1994; Jokipii and Jones 1996) has shown that if a coordinate is
ignorable (as in one- or two-dimensional hybrid simulations), cross-field
diffusion effects are suppressed, and since cross-field diffusion is an
essential part of injection and acceleration in shocks (certainly oblique ones)
{\it three-dimensional simulations} must be used to model shocks. No existing
computer is capable of running realistic three-dimensional plasma simulations
over dynamical ranges of energies appropriate to astrophysical applications. 
We emphasize that even though we model plane shocks and our scattering operator
is a gross simplification of the complex plasma processes taking place in
shocks, the operator is fully three-dimensional, includes cross-field
diffusion, and may well produce more realistic results than current one- or
two-dimension plasma simulations. In fact, comparisons between our model and
spacecraft observations of highly oblique interplanetary shocks (IPSs) (Baring
et al. 1995) suggest that this is the case.  The spacecraft observations clearly
show that highly oblique shocks inject and accelerate {\it thermal} particles,
a result we can model accurately (e.g. Figures~\ref{fig:sfspec}
and~\ref{fig:figacceff}), but one which, to our knowledge, all existing one-
and two-dimensional plasma simulations fail to show (e.g. Liewer, Goldstein,
and Omidi 1993; Liewer, Rath, and Goldstein 1995).

Virtually all analytic models of nonlinear shock acceleration have been
restricted to parallel shocks, however, Jones and Kang (1995) have recently
extended the cosmic ray diffusion-advection equation approach to oblique
geometry and produced impressive fits to the Ulysses observations mentioned
above (e.g. Baring et al. 1995).  However, all models based on the diffusion
approximation (i.e. the requirement that particle speeds be large compared to
flow speeds) are limited in their ability to treat thermal particles and must
use additional free parameters to model injection. For example,  the parallel
shock model of  Berezhko et al. (i.e. Berezhko, Yelshin, and Ksenofontov 1994;
Berezhko, Ksenofontov, and Yelshin 1995; Berezhko et al. 1995)  uses a source
term for monoenergetic injection at the gas subshock which is treated as a
discontinuity.  Here, a small fraction \teq{\epsilon} of incoming gas is
transferred to cosmic rays, the injected particles instantly obtaining a
superthermal momentum, \teq{p_{\rm inj}}.  Both \teq{\epsilon} and \teq{p_{\rm
inj}} are free parameters and the final results depend strongly on them.  The
main advantage of our model is that the Monte Carlo description is not
restricted to superthermal particles and injection is treated self-consistently
regardless of whether or not we assume that all particles obey
equation~(\ref{eq:mfp}).  Once some such scattering description is chosen, both
the injection rate and the effective injection momentum are fully determined,
as is the complete shock structure, by the Monte Carlo solution without any
additional parameters such as \teq{\epsilon}.  In fact, there is no `injection
momentum' in our solution since particles are drawn smoothly from the
background thermal gas.

The parameters which determine the injection and acceleration efficiency of
shocks are the obliquity, \teq{\Tbnone}, the strength of cross-field
diffusion, \teq{\eta}, the Mach numbers, \teq{\MS} and \teq{\MA}, and the size
of the shock system (i.e. \teq{|\dFEB |}).  These all influence the shock in
complex ways and there is no simple relationship between them.  In general, we
can state that the acceleration efficiency (i.e. the fraction of energy flux
which ends up in high energy particles) increases with:  (i) decreasing
\teq{\Tbnone},  (ii) decreasing \teq{\eta} (i.e. stronger scattering),  (iii)
increasing Mach number, and  (iv) increasing shock size. We have found that the
differences in the shock structure and acceleration efficiency resulting from
using either large-angle scattering (LAS) or pitch-angle diffusion (PAD) are
generally small (e.g. see Figure~\ref{fig:testspectra} for the test-particle
regime, and Figures~\ref{fig:specone} and~\ref{fig:sfspec} for full non-linear
results) for parameter regime we have investigated, namely \teq{\vht \ll c}, 
and can be neglected in the Bohm diffusion limit.  However, the differences
between the nonlinear results and the test-particle ones are very large (e.g.
Figure~\ref{fig:febspec}) except for high obliquities (i.e.
Figure~\ref{fig:sfspec}) or very low Mach numbers (Figure~\ref{fig:specone}),
where the acceleration  efficiency is low enough for the thermal gas to
dominate the non-thermal population dynamically, and the shock profiles are
very sharp.

For the first time, we have been able to calculate the absolute injection and
acceleration efficiency of non-linear oblique shocks without the use of an {\it
ad hoc} injection parameter. Our results (Figure~\ref{fig:figacceff}) show how
large differences in efficiency can occur as parameters change, however, we
have not yet explored the vast parameter regime oblique geometry opens up. Our
next step toward a more complete solution of the shock acceleration problem
will be a survey intended to quantify the differences the various parameters
make in the distribution function and overall efficiency. This will include
determining the effect of varying  equation~(\ref{eq:mfp}) and will yield
predictions for future spacecraft and plasma simulation results. The only way
to constrain our \teq{\eta} parameter is by comparing our results with direct
observations of shocks or with three-dimensional plasma simulations. Since, to
our knowledge, no three-dimensional simulation results showing significant
acceleration exist, we will concentrate on spacecraft observations as they
become available. As already mentioned, this work has begun with spectral
comparisons to Ulysses observations of nearby interplanetary shocks (i.e. those
not expected to encounter pick-up ions), and our preliminary comparisons with
data from Ulysses have already indicated that strong scattering accompanies
highly oblique IPSs, constraining \teq{\eta} to values smaller than about 10
(Baring et al. 1995). Our simulation produces particle distributions at
different distances upstream and downstream of the subshock, thereby providing
a wealth of model predictions for testing against observations.

Once we are successful in constraining our parameters with heliospheric shock
observations, the next step will be to apply our results to shocks with no \it
in situ \rm particle observations, such as the termination shock and supernova
remnant blast waves.  This is another useful aspect of our model and we expect
to be able to make predictions for the relative injection and acceleration
efficiency as a function of ionic composition for both thermal and pickup ions
at the termination shock, and to calculate how  efficiency varies around the
rim of a supernova remnant shock. Since relativistic particles are produced in
these shocks, our model will soon be generalized to include relativistic
particle energies.

\acknowledgements
The authors wish to thank B. Lembege, A. Mangeney, and S. Reynolds for
helpful comments.  Our code runs on massively parallel machines and
computing time for this project was provided by the Cray T3D's at CNRS
IDRIS and at C.E.A., Grenoble, France. This research was supported in
part by the NASA Space Physics Theory Program and DCE wishes to
thank the Service d'Astrophysique, C.E. Saclay, Observatoire de
Paris--Meudon, and CNET/CETP (Issy-les-Moulineaux) 
for hospitality during the time part of this work was performed.

\clearpage

\appendix
\section{APPENDIX}

This Appendix describes four technical aspects of the Monte Carlo simulation,
namely (1) our description of pitch angle diffusion, (2) how particles cross
the shock and grid points (actually planes of discontinuity of the flow and
field profiles) in the simulation, (3) the details of how the probability of
particle return from beyond the downstream simulation ``boundary'' is
determined, and (4) how we derive the form of the conservation of energy flux
in equation~(\ref{eq:enflux}).

\subsection{Pitch-Angle Diffusion}

To summarize our implementation of pitch-angle diffusion (PAD), which has been
given in detail in Ellison, Jones, and Reynolds (1990), we simulate small-angle
scattering effects by allowing the tip of the particle's fluid frame momentum
vector \teq{{\bf p}} to undergo a random walk on the surface of a sphere.  If
the particle originally had a pitch angle, \teq{\PitchBold = \arccos\{ 
{\bf p}\cdot {\bf B}/\vert {\bf p}\vert \,\vert{\bf B}\vert\} }, and after a
time \teq{\delta t} undergoes a small change in direction of magnitude
\teq{\delta\theta}, its new pitch angle, \teq{\PitchBnew}, is related to the
old by
\begin{equation}
   \cos\PitchBnew = \cos\PitchBold \cos{\delta \theta} -
   \sin{\PitchBold} \sin{\delta \theta} \cos{\phi}        \label{eq:pitchnew}
\end{equation}
where \teq{\phi} is the azimuthal angle of the momentum change 
\teq{\delta {\bf p}} measured relative to the plane defined by the original
momentum \teq{{\bf p}} and \teq{{\bf B}}. After each scattering, a new phase
angle around the magnetic field, \teq{\phiBnew}, is determined from the old
phase angle, \teq{\phiBold}, by
\begin{equation}
   \phiBnew = \phiBold + \arcsin \left[ 
   \dover{ \sin{\phi}\, \sin{\delta \theta} }{ \sin{\PitchBnew} } \right] 
  \label{eq:phasenew}
\end{equation}
\teq{\delta \theta} is randomly chosen from a uniform distribution between
\teq{0} and \teq{\delta \theta_{\rm max}}, and \teq{\phi} is randomly chosen
from a uniform distribution between \teq{-\pi} and \teq{\pi}, so that the tip
of the momentum vector walks randomly over the surface of a sphere of radius 
\teq{p=\vert {\bf p}\vert}.

If the time required to accumulate deflections of the order of \teq{90^\circ} 
is identified with the collision time, \teq{t_c}, using a diffusion analysis,
the relation between \teq{\delta \theta_{\rm max}} and the mean free path
\teq{\lambda} was shown by Ellison, Jones, and Reynolds (1990) to be
\begin{equation}
   \delta \theta_{\rm max} = \sqrt{ 6\dover{\delta t}{t_{\rm c}} }
\end{equation}
where \teq{t_{\rm c}=\lambda/v}.  Pitch angle diffusion is then defined by the
regime \teq{\delta t\ll t_{\rm c}}.  Clearly, using this approach implies a
magnetic fluctuation correlation length  smaller than the particle gyroradius;
this method then becomes an approximation that is nevertheless still very
convenient for implementation in Monte Carlo simulations.  Note that in the
limit of \teq{\delta \theta_{\rm max}\to 2\pi}, this prescription of PAD
becomes comparable to our scheme for large angle scattering (see Ellison,
Jones, and Reynolds 1990).

Equations~(\ref{eq:pitchnew}) and~(\ref{eq:phasenew}) then can be used to
determine the coordinates of the new gyrocenter 
\begin{eqnarray}
   x_{\rm gc}  = & x \ - & \rg\cos\PitchBnew \cos{\left (\phiBnew - 
     \dover{\pi}{2} \right )} \sin{\Tbn}\nonumber\\
   y_{\rm gc}  = & y \ + & \rg\sin\PitchBnew \sin{\left (\phiBnew - 
     \dover{\pi}{2} \right )}\\
   z_{\rm gc}  = & z \ + & \rg\sin\PitchBnew \cos{\left (\phiBnew - 
     \dover{\pi}{2} \right )} \cos{\Tbn}\nonumber
\end{eqnarray}
where \teq{(x,y,z)} is the position of the particle when it scatters.  The
phase offset of \teq{\pi /2} represents the differences between position and
momentum vector phases.  The gyroradius \teq{r_{\rm g}} remains unchanged in
the PAD event (true also for large angle scattering) since \teq{\vert {\bf
B}\vert} is fixed at the point of scattering and our assumption of elastic
scattering leaves the magnitude of the momentum unchanged in the fluid frame. 
However, in contrast to the LAS case where the particle's momentum vector is
only updated after \teq{t_{\rm c}} on average, the momentum vector is updated
after every \teq{\delta t} for PAD.

\subsection{Shock or grid zone boundary crossing}

When a particle crosses a grid zone boundary (there is no distinction in our
code between the shock and any other grid boundary), its orbit is changed
because the magnetic field changes direction and magnitude.  The new values of
the particle's pitch angle are obtained from the assumption that the momentum
in the HT frame remains unchanged at the zone boundary; this follows from of
the absence of drift electric fields in this frame.  This method does not
require that the magnetic moment be conserved; differences between gyrohelix
computations at a flow interface and the adiabatic approximation are discussed
by Terasawa (1979).

\placefigure{fig:gdcross}          

The detailed calculation  (see also Decker 1988; Begelman and Kirk 1990;
Ostrowski 1991; and Takahara and Terasawa 1991)  is as follows (the geometry is
illustrated in Figure~\ref{fig:gdcross}).  The component of the old momentum 
(i.e. the momentum before crossing the grid zone boundary) in the $y$-direction
is given by
\begin{equation}
  p_{\rm y} \; =\;
  p_\perp^{\hbox{\sixrm O}} \sin{\phi_{\hbox{\sixrm B}}^{\hbox{\sixrm O}}}
\end{equation}
and the component along the $z^\prime$-direction is
\begin{equation}
  p_{\rm z'} \; =\; p_\perp^{\hbox{\sixrm O}} 
  \cos{\phi_{\hbox{\sixrm B}}^{\hbox{\sixrm O}}}
\end{equation}
where $p_\perp^{\hbox{\sixrm O}}$ is the component of momentum along  ${\bf
B_{\rm i}}$, and $z'$ is perpendicular to the  $y$-${\bf B_{\rm i}}$ plane.
Note that all momenta in this section are measured in the HT frame. The
component of momentum along the $z''$-direction (i.e. the axis perpendicular to
the $y$-${\bf B_{\rm i+1}}$ plane), is given by
\begin{equation}
  p_{\rm z^{\prime\prime}} \; =\;
  p_{\rm z^\prime} \cos{\Delta\Theta_{\rm Bn}} -
  p_{\hbox{\sixrm B}}^{\hbox{\sixrm O}} \sin{\Delta\Theta_{\rm Bn}}
\end{equation}
where $\Delta\Theta_{\rm Bn}$ is the difference  in $\Tbn$ across the grid zone
boundary. The total momentum perpendicular to the new magnetic field direction,
${\bf B_{\rm i+1}}$, is
\begin{equation}
  p_\perp^{\hbox{\sixrm N}} \; =\;
  \sqrt{p_{\rm y}^2 + p_{\rm z^{\prime\prime}}^2}
\end{equation}
and the new momentum parallel to the new magnetic field direction is
\begin{equation}
  p_{\hbox{\sixrm B}}^{\hbox{\sixrm N}} \; =\;
  p_{\hbox{\sixrm B}}^{\hbox{\sixrm O}} \cos{\Delta\Theta_{\rm Bn}} +
  p_{\rm z^\prime} \sin{\Delta\Theta_{\rm Bn}}
\end{equation}
Finally, the new phase around ${\bf B_{\rm i+1}}$ is given by
\begin{equation}
\phiBnew \; =\; 
\arctan{ {p_{\rm y}} \over p_{\rm z^{\prime\prime}}}
\end{equation}

\subsection{The Probability of Return Calculation}

The details of how particle return from the far side of the downstream return
boundary (DRB) is effected are presented here; such return boundaries are found
not only in our Monte Carlo technique, but also in hybrid plasma simulations 
(e.g Bennett and Ellison 1995).  Assume a uniform flow with a component of
velocity in the positive \teq{x}-direction of  \teq{u_{\rm x2}} and assume that
particles in the local fluid frame are isotropic and of speed \teq{\vf}. 
Quantities denoted by subscript F are measured in this fluid (plasma) frame.  
The flux of particles crossing a \teq{y}-\teq{z} plane that is parallel
to, and downstream of, the subshock interface is proportional to \teq{\vxsk},
the \teq{x}-component of particle speed in \it any \rm frame in which the
shock is at rest.  Here \teq{\vxsk >0} (\teq{<0}) for transmissions to the
downstream (upstream) side of the DRB.  

The probability that particles return to the DRB after crossing it from the
upstream side is therefore simply (e.g. see Jones and Ellison 1991) the ratio
of the flux of particles moving upstream of the DRB to the flux of particles
moving to the downstream side of this plane.  Clearly \teq{0<\vxsk < \vf
+u_{\rm x2}} defines downstream crossings of the DRB, while \teq{-\vf +u_{\rm
x2}<\vxsk < 0} prescribes upstream crossings.   We confine the discussion to
cases with \teq{\vf > u_{\rm x2}}, and further specialize hereafter to the
specific shock rest frame where the flow has zero component of velocity in the
plane of the DRB.  Integrating over the angle of the particle velocity relative
to the shock normal, or alternatively over \teq{\vxsk}, the probability of
return \teq{P_{\rm ret}} to the DRB for isotropic particles of speed \teq{\vf}
(in the fluid frame) is
\begin{equation}
  P_{\rm ret}\; =\; 
  \left | \dover{\int_{-\vf +u_{\rm x2}}^0 \vxsk d\vxsk \vphantom{\Biggl)}}{
  \int_0^{\vf + u_{\rm x2}} \vxsk d\vxsk \vphantom{\Biggl)}} \right | 
  \; =\; {\left ( \dover{\vf -u_{\rm x2}}{\vf +u_{\rm x2} } \right )}^2 \quad .
 \label{eq:probreturn}
\end{equation}
This expression is valid for any shock obliquity and is relativistically
correct (Peacock 1981). It applies to all \teq{\vf\geq u_{\rm x2}}
(for \teq{\vf <u_{\rm x2}},
\teq{P_{\rm ret}=0}), and the only requirements for its validity are
that the particles be isotropic in the local fluid frame and that they do not
change speed in the region to the right of the return plane.  To ensure
isotropy, we only apply equation~(\ref{eq:probreturn}) after particles have
scattered at least once in the downstream region when LAS is used, or that
particles have diffused through \teq{90^\circ} in the downstream region when PAD
is used.  The decision for return, or otherwise, is made via a random number
generator.

Now, consider only those particles which return back across the return plane. 
Both their pitch angle relative to the magnetic field, \teq{\PitchB =\arccos\{
{\bf p}\cdot { \bf B}/\vert {\bf p}\vert \,\vert{\bf B}\vert\}}, and their phase
around the magnetic field, \teq{\PhaseB} must be determined in the fluid frame.
Determining these requires knowledge of \teq{\vxsk} for the returning
particles, which can be computed by again noting that the flux of
particles returning across the return plane (moving in the negative
\teq{x}-direction) is proportional to \teq{\vxsk}.  This means that the number
of particles returning with \teq{\vxsk} between \teq{\vxsk} and \teq{\vxsk +
d\vxsk} is proportional to \teq{\vxsk d\vxsk}.  So, for a particular \teq{\vf},
returning particles are drawn from a distribution such that
\begin{equation}
   \dover{2}{(\vf - u_{\rm x2})^2} \int^0_{\vxsk} {\vxsk' d\vxsk'}
   \; =\; \int_0^{N_{\rm R}} {dN_{\rm R}} \; =\; N_{\rm R}
\end{equation}
where \teq{N_{\rm R}} is a random number uniformly distributed between 0 and 1.
Therefore,
\begin{equation}
   \vxsk \; =\; \sqrt{N_{\rm R}} (-\vf + u_{\rm x2}) \quad ,
\end{equation}
and from this the \teq{x}-component of speed in the plasma frame
\begin{equation}
   \vxf \; =\; \vxsk - u_{\rm x2} \; =\; -\vf\sqrt{N_{\rm R}}-u_{\rm x2}\,
    (1-\sqrt{N_{\rm R}})
\end{equation}
can be obtained.  Choosing a series of random numbers, \teq{N_{\rm R}}, between
0 and 1 gives the proper distribution of returning particles.

\placefigure{fig:probretone}          

To determine \teq{\PitchB} and \teq{\PhaseB} in the fluid frame, we assume that
particles will return distributed symmetrically around the \teq{x}-axis  in the
fluid frame, a consequence of isotropy.  This symmetry appears also in our
special shock frame (see Figure~\ref{fig:probretone}), where the flow is
orthogonal to the DRB, since phases are preserved in velocity transformations
orthogonal to this plane.  The velocity vector of a returning particle will
make an angle \teq{\theta_{\rm x}} in the fluid frame with the \teq{x}-axis,
given by \teq{\cos\theta_{\rm x}=\vxf/\vf =(\vxsk - u_{\rm x2})/\vf}.  If it
also has an azimuthal angle about the \teq{x}-axis of \teq{\varphi_{\rm x}}, 
chosen randomly between 0 and \teq{2\pi}, then from
Figure~\ref{fig:probretone}, using spherical triangles, the values of the fluid
frame pitch angle \teq{\PitchB} and phase \teq{\PhaseB} can be expressed in
terms of \teq{\theta_{\rm x}}, \teq{\varphi_{\rm x}} and \teq{\Tbntwo}. 
Subsequently, the returning particle is placed at the  downstream return plane
and, using the new fluid frame phase and pitch angle (and also a new position
for the guiding center), propagated upstream by transforming to the de
Hoffmann-Teller frame.  The statistical prescription in this subsection
guarantees that our code effectively simulates an infinite region to the
downstream side of the probability of return plane.

\subsection{The energy flux conservation equation}

While the pressure terms in the momentum fluxes in equations~(\ref{eq:xmoflux})
and~(\ref{eq:zmoflux}) are elementary to write down, derivation of the forms for
the corresponding terms in the energy flux in equation~(\ref{eq:enflux}) involve
some subtleties.  As outlined Section~3.1, we restrict the analysis to 
gyrotropic particle distributions in local fluid frames in the interest of
simplicity; we believe that such an approximation is quite good, and certainly
better than the assumption of isotropy that is ubiquitous in flux conservation
equation usage in the literature.

In a local fluid frame somewhere in the shock environs, consider coordinate 
axes oriented so that the $x$-direction is aligned with the local magnetic
field, but such that a single rotation about the $y$-axis produces an 
identical orientation to the system depicted in
Figure~\ref{fig:shock}. 
In this coordinate
system, a gyrotropic plasma has a diagonal pressure tensor, namely \teq{P_{\rm
xx}\equiv P_{\parallel}} and \teq{P_{\rm yy}=P_{\rm zz}\equiv  P_{\perp}} with
\teq{P_{ij}=0} otherwise.  \teq{P_{\parallel}} and \teq{P_{\perp}} are
components of pressure parallel to and orthogonal to the ambient field, and for
a thermal plasma are related to analogous temperature components by two
equations of state.  Rotating the axes into alignment with the system in
Figure~\ref{fig:shock} 
yields a non-diagonal pressure tensor \teq{\cal P} due to mixing of
the components: 
\begin{equation}
{\cal P}\; =\; {\cal R}  
      \left( \begin{array}{ccc}
             P_{\parallel} & 0 & 0 \\
                     0 & P_{\perp} & 0 \\
                         0 & 0 & P_{\perp} \end{array} \right) 
{\cal R}^{-1} \quad , \quad {\cal R} \; =\;
  \left( \begin{array}{ccc} \cos\Tbn & 0 & \sin\Tbn  \\
                                   0 & 1 & 0 \\
                     -\sin\Tbn & 0 & \cos\Tbn  \end{array} \right)\;\; ,
\end{equation}
where \teq{{\cal R}} is the rotation matrix.  This gives a specific form
for the pressure tensor of  
\begin{equation}
{\cal P} \; =\;
\left( \begin{array}{ccc}
    P_{\parallel} \cos^2\Tbn + P_{\perp}\sin^2\Tbn & 0 & 
       (P_{\perp}-P_{\parallel}) \sin\Tbn\cos\Tbn  \\
                   0 & P_{\perp} & 0 \\
       (P_{\perp}-P_{\parallel}) \sin\Tbn\cos\Tbn & 
    0 & P_{\parallel} \cos^2\Tbn + P_{\perp}\sin^2\Tbn \end{array} \right)
    \quad .   \label{eq:ptensor}
\end{equation}
Since pressure represents the spread of velocities about the mean speed, the
pressure tensor is invariant under bulk velocity transformations.  Hence
it follows that equation~(\ref{eq:ptensor}) defines the pressure in the normal
incidence frame, and therefore is directly applicable to the flux conservation
considerations.  The above coordinate rotation therefore yields the
relationships
\begin{eqnarray}
   P_{\parallel} & = & P_{\rm xx} + P_{\rm xz} \tan\Tbn \nonumber\\
   P_{\perp} & = & P_{\rm xx} + P_{\rm xz} \cot\Tbn\quad ,
\end{eqnarray} 
which define the components of the pressure tensor on the fluid frame in
terms of normal incidence frame tensor components that can be simply
determined using the structure of our simulation.  At this point it 
becomes apparent that the assumption of gyrotropy in the fluid frame is
indeed expedient, since it enables complete specification of the fluid
frame pressure tensor using only flux quantities measured in the NIF in the 
$x$-direction: generalizing from the gyrotropic approximation would require
construction of coordinate grids in the other directions, thereby complicating
the simulation immensely, with only marginal gain in physical accuracy.

The energy flux equation can now be simply constructed from the formalism on
p.~56 of Boyd and Sanderson (1969).  The convective contribution to the energy
flux is simply \teq{P_{\rm xx}u_x+P_{\rm zz}u_z}.  The thermal-type (i.e.
velocity spread) term is usually written in the form \teq{\rho\kB T\,
u_x/(\gamma -1)} for temperature \teq{T}, where \teq{\kB} is Boltzmann's
constant, and \teq{\gamma} is the ratio of specific heats.  For the non-thermal
application here, prescribing the temperature is inappropriate, so we make use
an equation of state \teq{\rho\kB T ={\rm Tr}\{ {\cal P} \}=P_{\parallel}
+ 2P_{\perp}} in order to generalize to non-thermal situations by converting
to pressure formalism.  It follows that 
\begin{equation}
   \dover{u_x}{3(\gamma -1)} (P_{\parallel}+2P_{\perp}) \; =\;
   \dover{u_x}{(\gamma -1)} \biggl\lbrack\, P_{\rm xx} + \dover{1}{3}\, 
   P_{\rm xz} \bigl( \tan\Tbn + 2\cot\Tbn  \bigr) \,\biggr\rbrack\quad .
 \label{eq:thermp}
\end{equation}
is the non-convective contribution to the energy flux in the $x$-direction
that results from particle pressure.  This applies to both the non-relativistic
gases considered in this paper (with \teq{\gamma =5/3}), and also more general
cases with different equations of state (i.e. \teq{4/3 <\gamma <5/3}),
including relativistic plasmas.  Using the magnetic field identity
\teq{B_z/B_x=\tan\Tbn} then results in the form given in
equation~(\ref{eq:enflux}).

\clearpage

\clearpage


\figureout{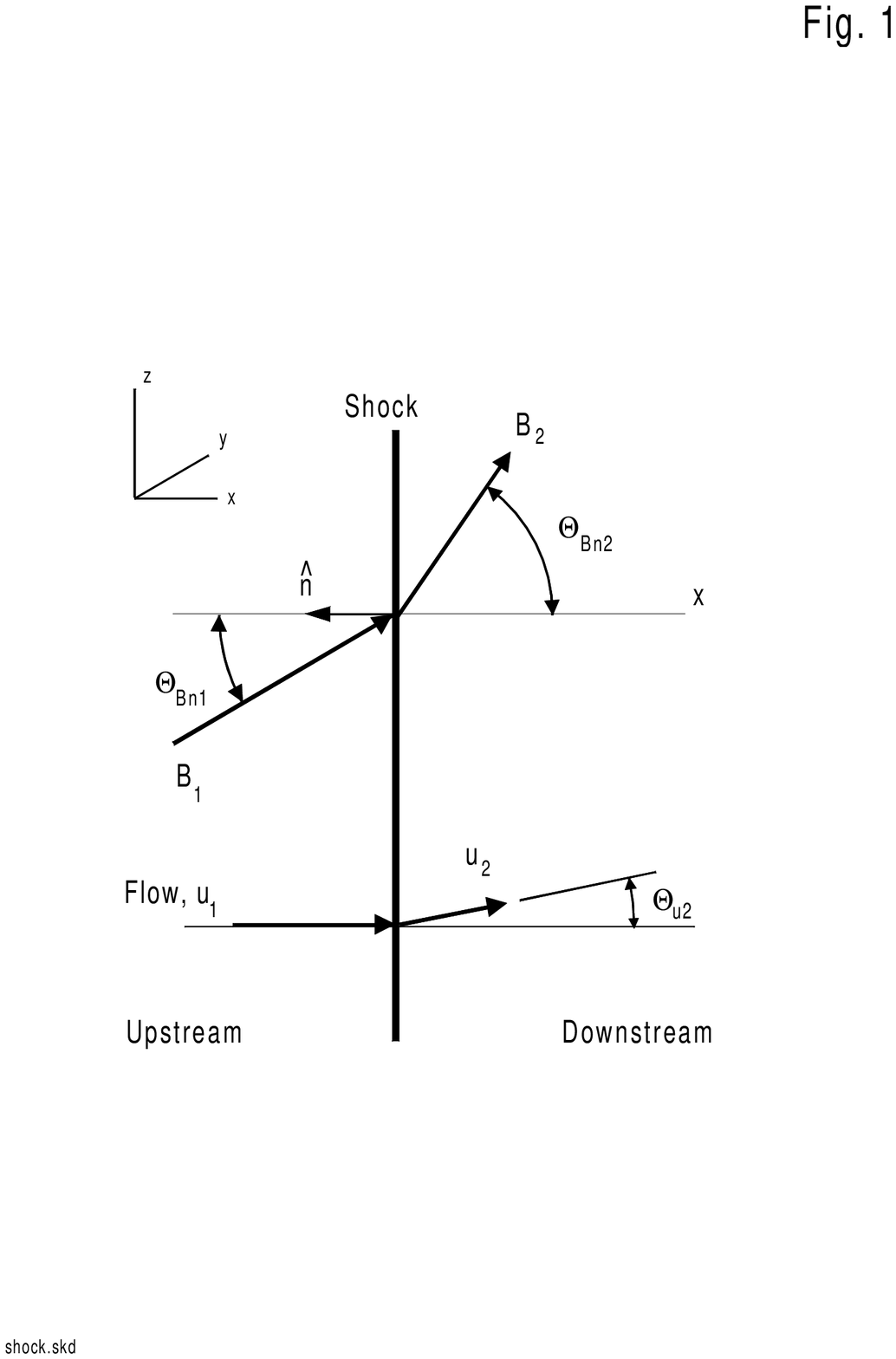}{
  The geometry of a discontinuous plane shock in the normal incidence frame
  (NIF), the frame where particle distributions and fluxes are output from the
  simulation.  The shock lies in the \teq{y}-\teq{z} plane and all quantities
  vary only in \teq{x}.  The geometry of the `smooth' shock is similar except
  that the magnetic field and flow make many small kinks instead of one large
  one, i.e. the profiles consist of a sequence of connected pieces like this
  depiction.     \label{fig:shock}
}

\figureout{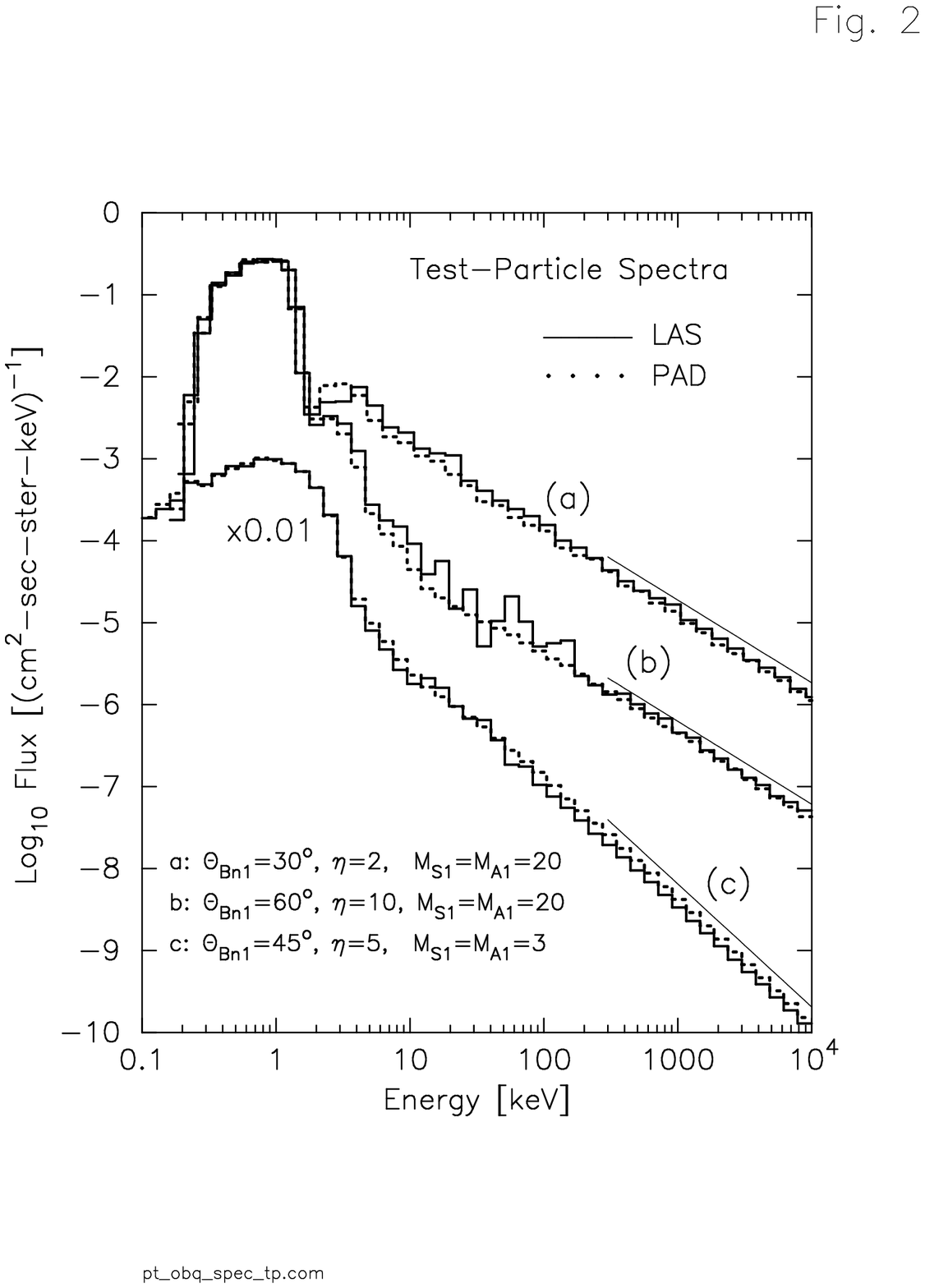}{
  Test-particle omni-directional, distribution functions measured several
  \teq{\lambda_0} downstream from a discontinuous shock in the normal incidence
  frame.  All spectra here and elsewhere are normalized to one particle per
  square cm per second injected far upstream.  Each pair of histograms has
  obliquity, Mach numbers and \teq{\eta} as indicated according to the labels
  (a), (b) and (c). The solid lines are results using large-angle scattering,
  while the dotted lines are for pitch-angle diffusion.  At energies where
  \teq{v\gg u_{\rm x1}/\cos\Tbnone} applies, all spectra attain the canonical
  Fermi power-law (solid lines of arbitrary normalization). Note that the
  spectra labeled (c) are multiplied by 0.01 for clarity.  The similarities
  for the two modes of scattering are striking.        \label{fig:testspectra}
}

\figureout{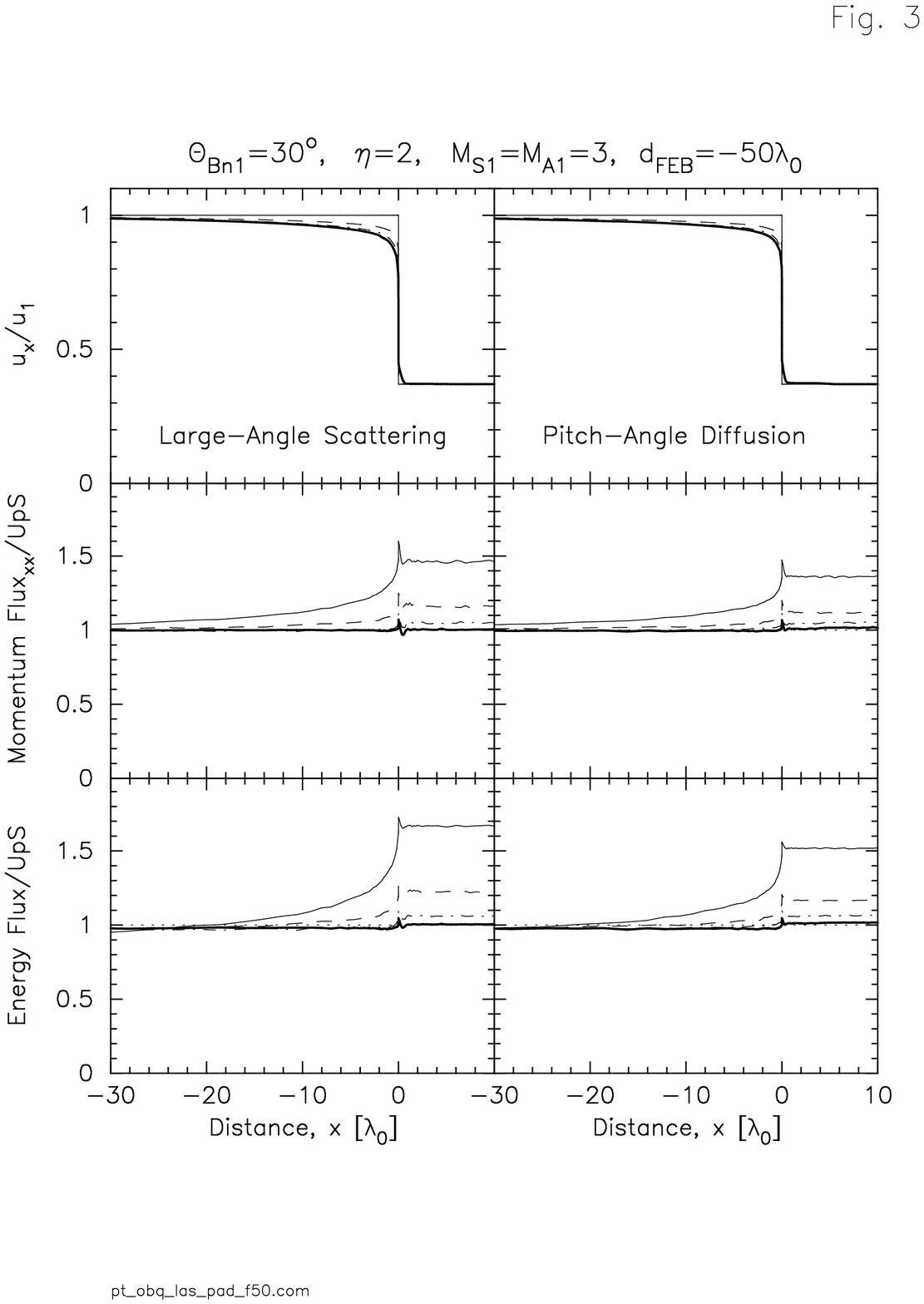}{
  Flow and flux profiles for quantities in the \teq{x}-direction for a weak
  shock with parameters \teq{\MA=3}, \teq{\MS=3}, \teq{\Tbnone=30\deg},
  \teq{\dFEB=-50\lambda_0}, \teq{\eta=2}, and a self-consistently determined
  compression ratio, \teq{r=2.7}.   The three left panels show the
  \teq{x}-component of the flow speed, the flux \teq{\Fpxx} of the
  $x$-component of 
  momentum, and the energy flux for large-angle scattering.  All quantities are
  measured in units of the far upstream values (UpS).  The three right panels
  show the same quantities for pitch-angle diffusion.  In  all cases, four
  iterations are shown; the first with a light solid line, the second with a
  dashed line, the third with a dot-dashed line, and the fourth with a heavy
  solid line (the flat dotted line indicates upstream values).  After four
  iterations, all quantities remain the same except for statistical variations.
  The momentum and energy fluxes are conserved everywhere 
  to within 10\%. Note that previous iterations (not shown) were
  done to determine the compression ratio.  The different scattering modes
  produce identical results within statistics.    \label{fig:twoprofiles}
}

\figureout{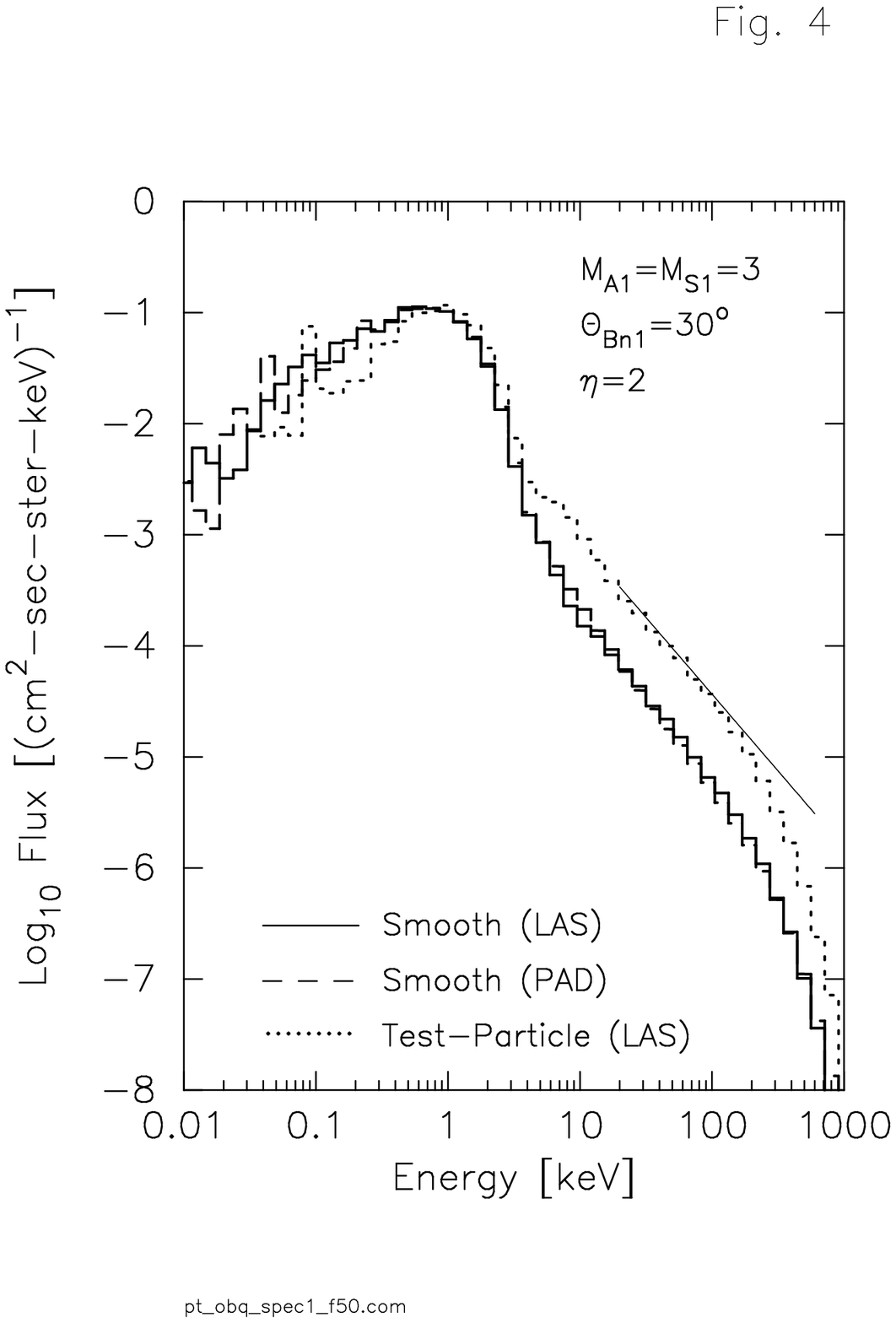}{
  Distribution functions measured downstream from the shock in the shock (or
  normal incident) frame obtained from the profiles shown in
  Figure~\ref{fig:twoprofiles}.   The turnover at \teq{E\sim 100} keV is the
  result of particles leaving at the upstream FEB.  Apart from statistics, there
  are no discernible differences between the large-angle scattering and
  pitch-angle diffusion scattering modes.  For comparison, the dotted line shows
  the test-particle spectrum obtained for the same shock parameters and the
  light solid line is the expected Fermi power-law (arbitrary normalization) 
  for \teq{r=2.7}.      \label{fig:specone} 
}

\figureout{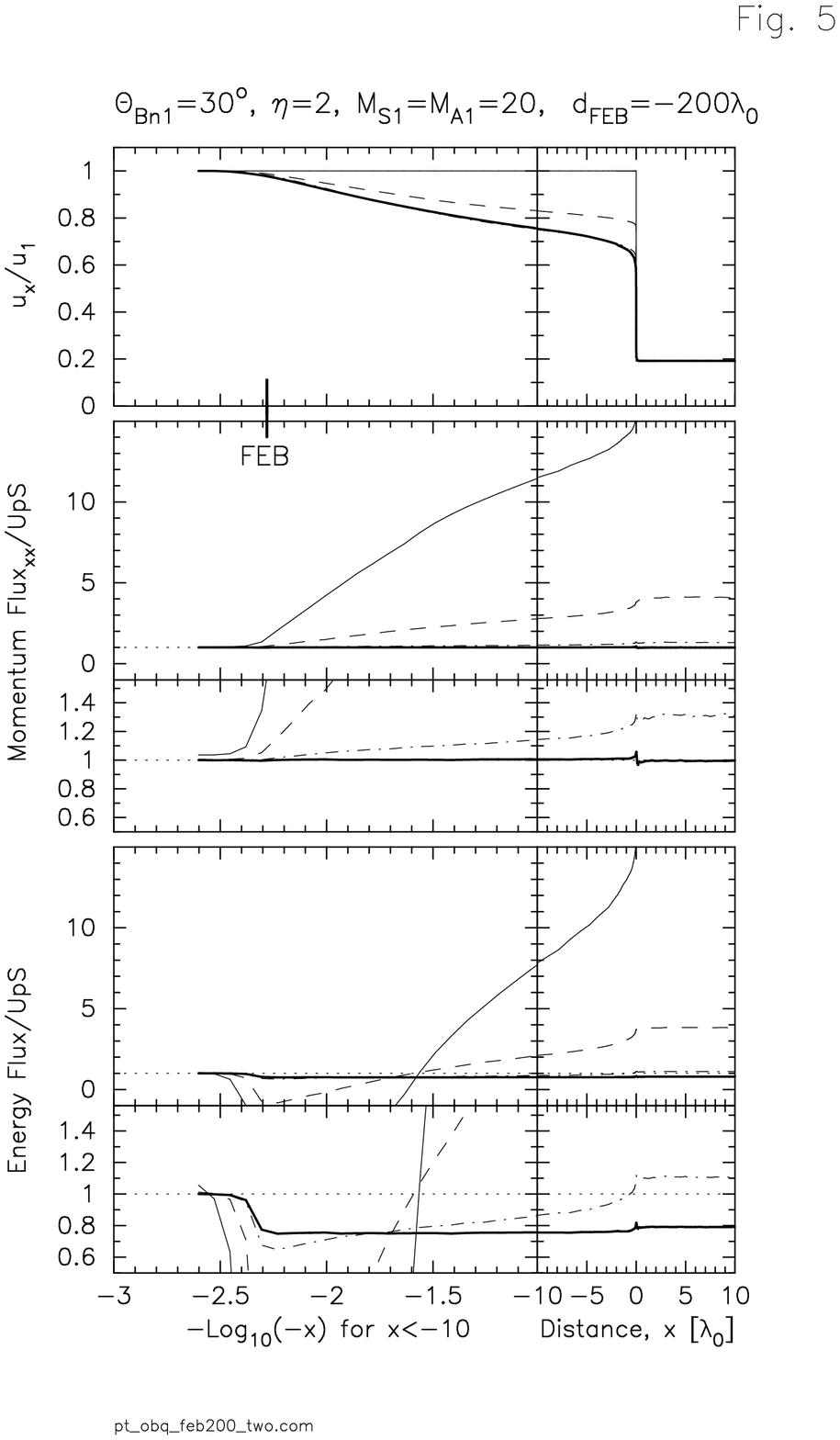}{
  The top panel shows the \teq{x}-component of flow speed versus \teq{x}. The 
  next two panels show the flux \teq{\Fpxx} of the $x$-component of momentum,
  and the bottom two panels show the energy flux. The shock parameters are:  
  \teq{\MS=\MA=20}, \teq{\dFEB=-200\lambda_0}, \teq{u_1=500}km~sec$^{-1}$,  
  \teq{n_1=1}cm$^{-3}$, \teq{\Tbnone=30\deg}, and \teq{\eta=2}. In all cases,
  four iterations are shown the first, second, third, and fourth iterations
  are shown by light solid lines, dashed lines, dash-dot lines, and heavy
  solid lines, respectively, and the horizontal scale is logarithmic for
  \teq{x<-10\lambda_0} and linear for \teq{x>-\lambda_0}.  After four
  iterations, no further changes occur in the profiles and momentum and energy
  are conserved across the shock once the escaping energy flux 
  (\teq{Q'_{\rm esc}/F'_{\rm en1}\simeq 0.17}) is accounted for.  Note that
  previous iterations were performed to obtain the self-consistent compression
  ratio, \teq{r\simeq 5}.       \label{fig:febprofile}
}

\figureout{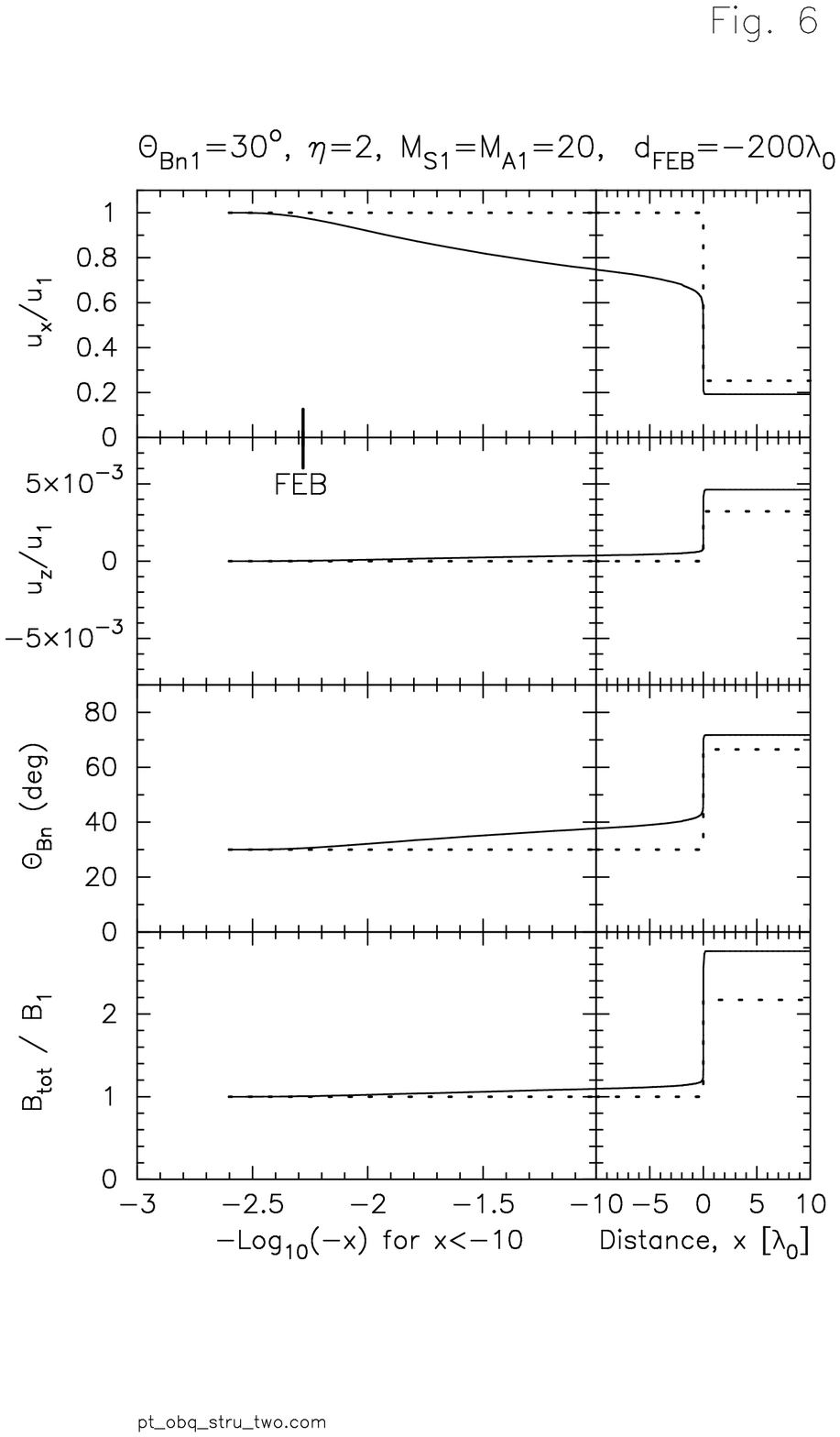}{
  Complete shock structure (solid lines) for the example shown in
  Figure~\ref{fig:febprofile} again with a composite distance scale. The dotted
  lines show the discontinuous shock structure with the initial R-H compression
  ratio, \teq{r=3.96}.  The final shock has a overall compression ratio of
  \teq{r\simeq 5}.                      \label{fig:febstruct}
}

\figureout{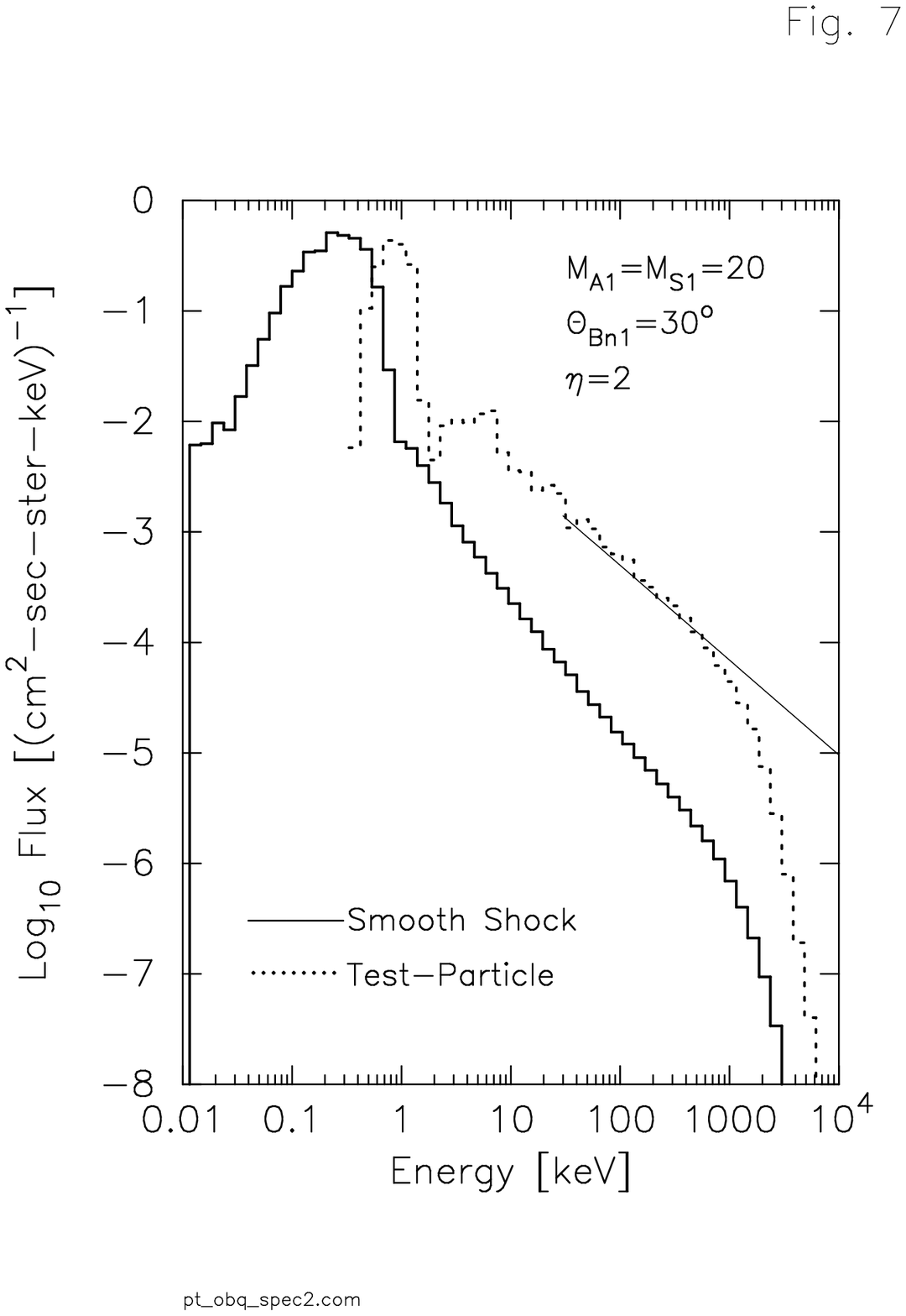}{
  The downstream, shock frame distribution functions for the smooth shock
  obtained in Figure~\ref{fig:febprofile} (heavy solid line) and a test-particle
  shock (dotted line) with the same parameters.  Note the shift of the thermal
  peak to lower energy which occurs in the smooth shock.  The light solid line
  shows the test-particle, power-law slope expected from Fermi acceleration
  for a shock with \teq{r=5}.  This is obtained by the discontinuous shock
  solution before the falloff produced by the FEB, but not by the smooth shock
  solution.             \label{fig:febspec}
}

\figureout{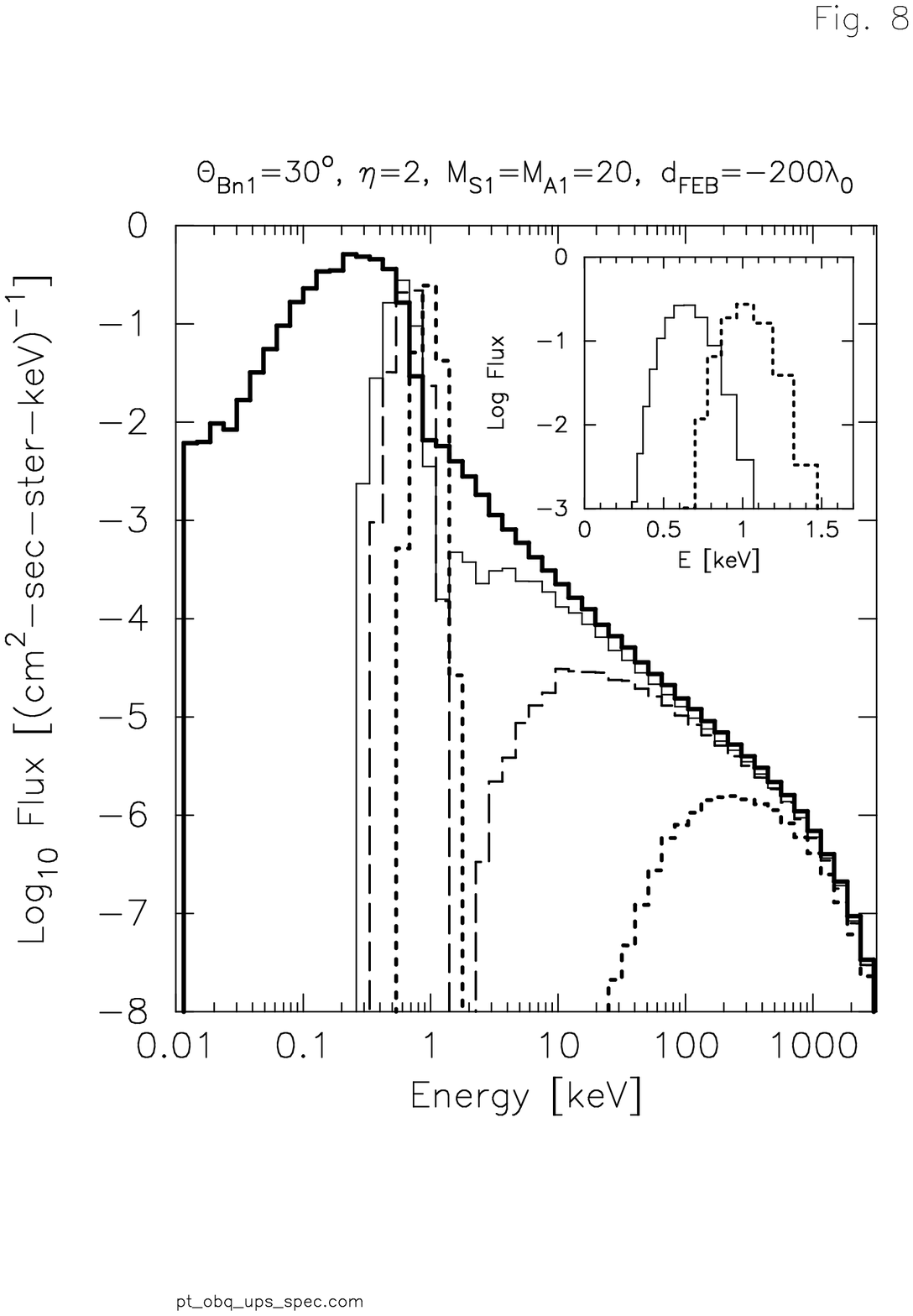}{
  Shock frame distribution functions calculated at  \teq{x= -50 \lambda_0}
  (dotted line), \teq{x= -4 \lambda_0} (dashed line), \teq{x=-0.5 \lambda_0}
  (light solid line), and  \teq{x=+\lambda_0} (heavy solid line).  The insert
  shows the thermal peaks for the \teq{x=-50 \lambda_0} and \teq{x=-0.5
  \lambda_0} cases on a linear energy scale.   The \teq{-4\lambda_0} thermal
  peak lies between these two.            \label{fig:upsspec}
}

\figureout{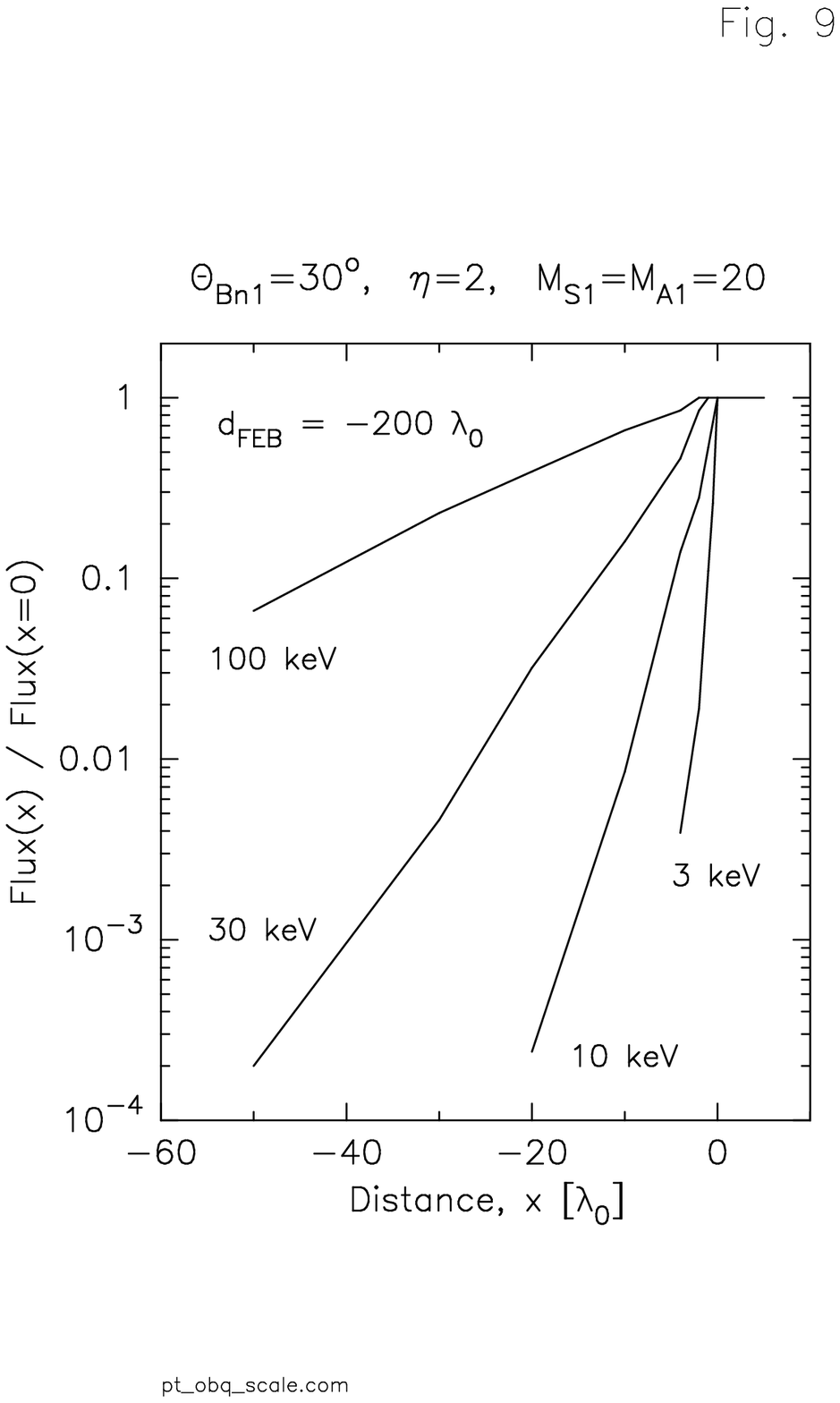}{
  Ratio of the omni-directional flux, \teq{dJ/dE}, at position \teq{x} to the
  flux at \teq{x=0} for a given energy  versus distance upstream from the shock.
  These define the scale heights for the several energies  indicated on the
  figure.        \label{fig:upsscale}
}

\figureout{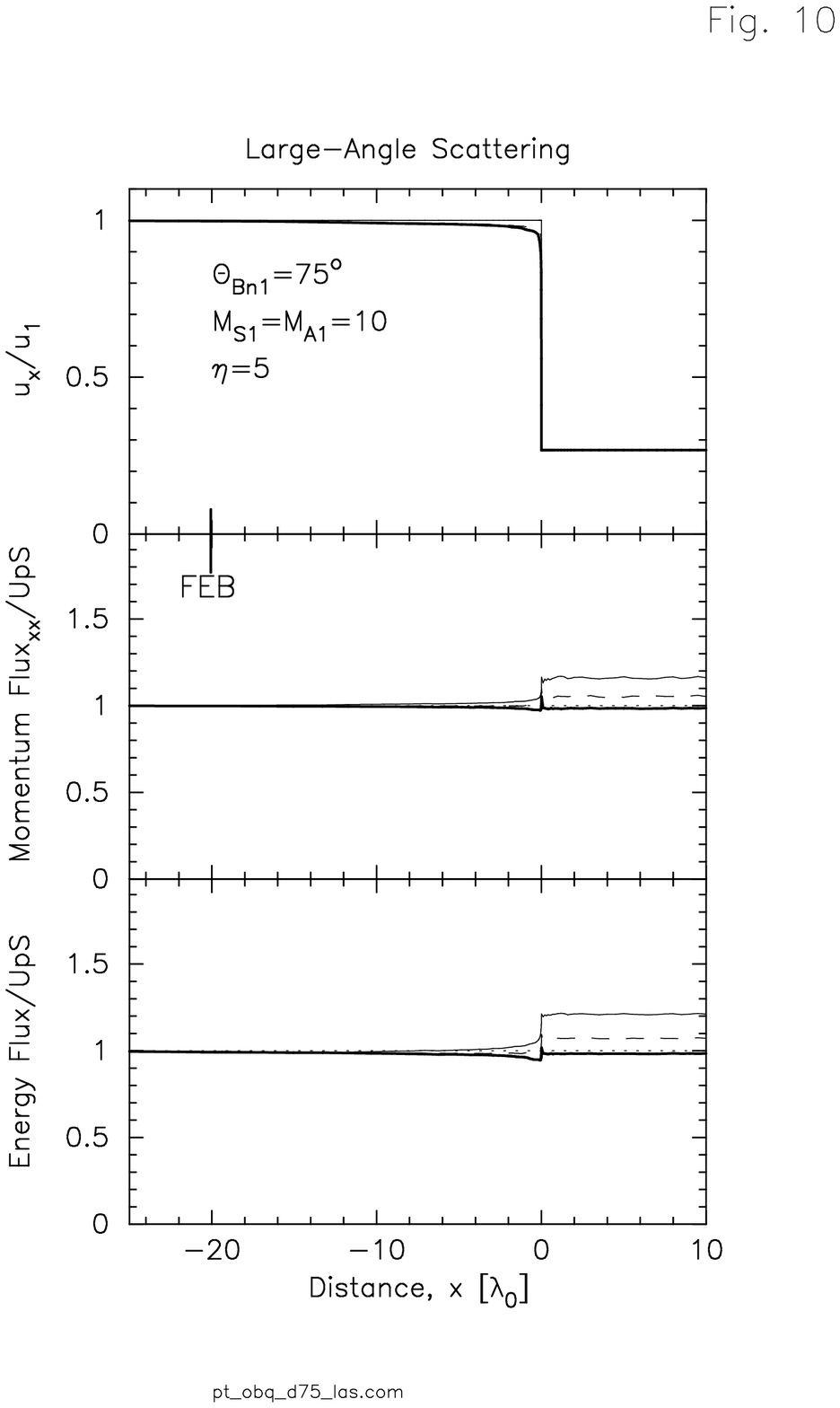}{
  The three panels show  the \teq{x}-component of the flow speed, the flux
  \teq{\Fpxx} of the $x$-component of momentum, and the energy flux for a shock
  with \teq{\Tbnone=75\deg}, \teq{\MS =\MA =10}, \teq{\dFEB=-20 \lambda_0},
  \teq{\eta=5}, and large angle scattering. The profile for the PAD case is
  essentially the same and is not shown. In all panels, the first, second,
  third, and fourth iterations are shown by light solid lines, dashed lines,
  dash-dot lines, and heavy solid lines, respectively.  Comparatively little
  acceleration takes place and the final shock profile is very nearly
  discontinuous with the R-H compression ratio, \teq{r=3.74}.            
    \label{fig:sfprofile}
}

\figureout{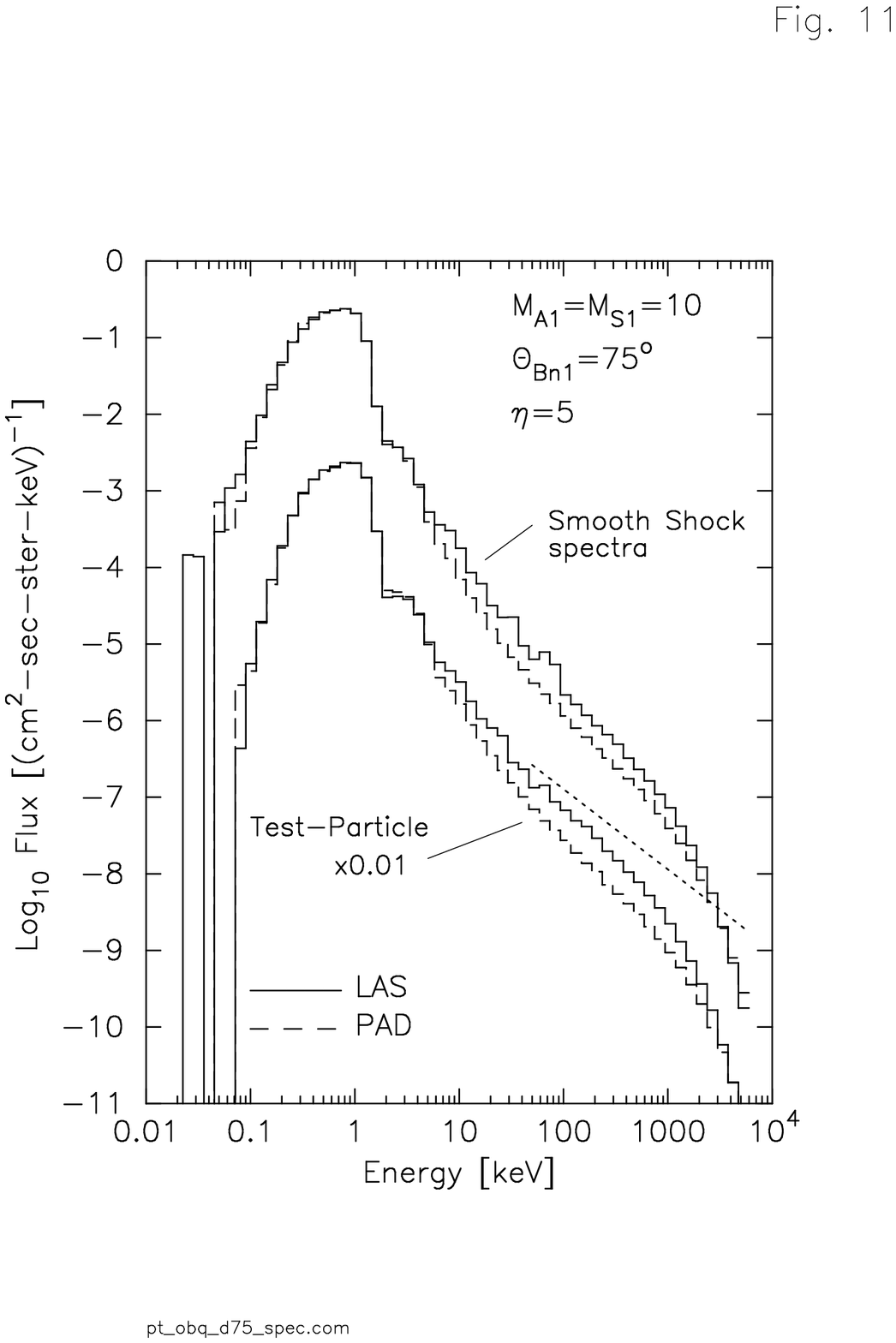}{
  Distribution functions for a highly oblique shock (\teq{\Tbnone=75\deg}). The
  top two curves are smooth shock solutions using LAS (solid line) and PAD
  (dashed line).  The lower two curves are test-particle results.  The straight
  dotted line shows the slope expected from the standard Fermi power-law. Unlike
  our examples at lower obliquities, clear differences exist depending on the
  type of scattering although these are not great enough to produce noticeable
  differences in the shock profile.         \label{fig:sfspec}
}

\figureout{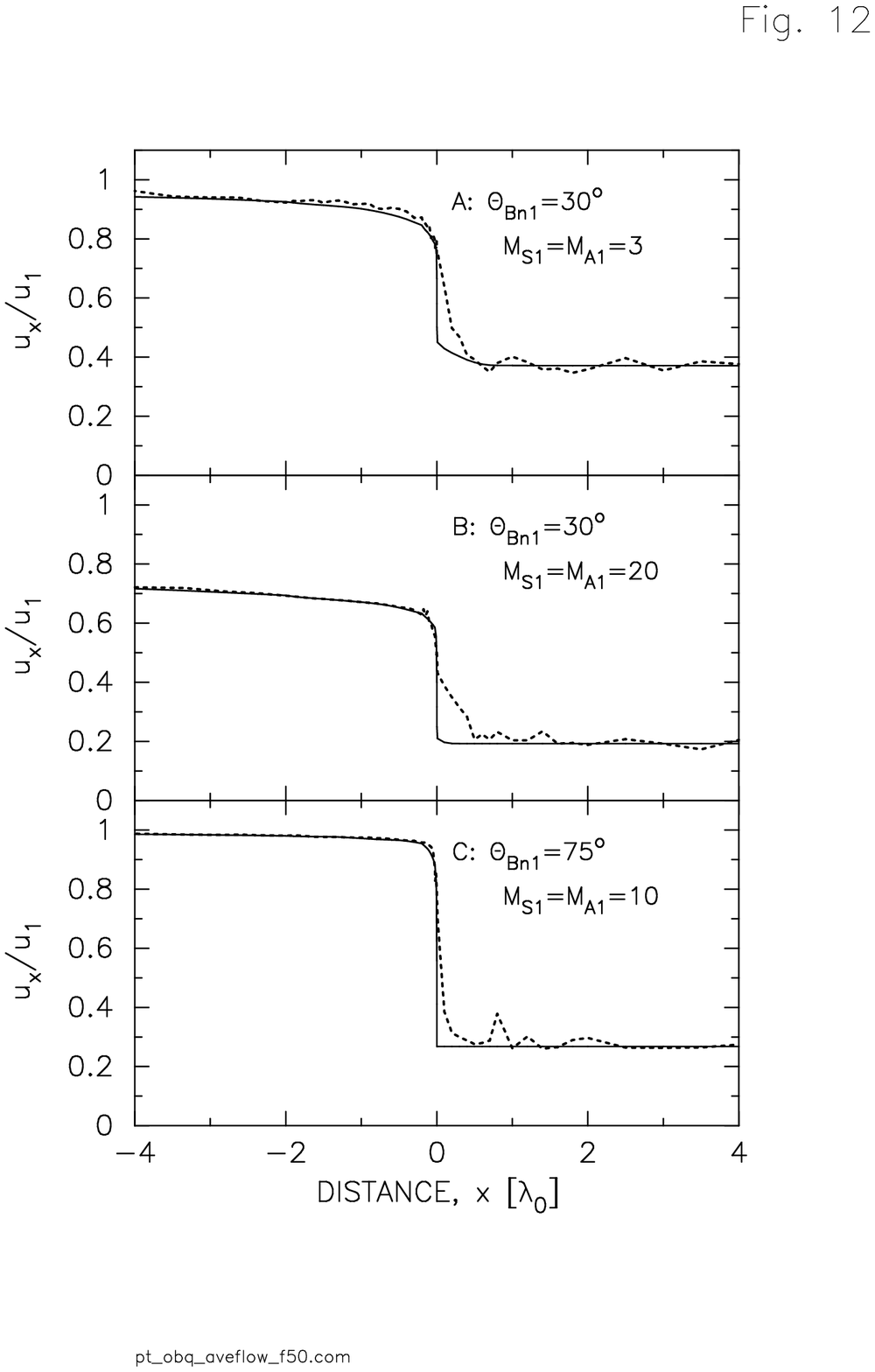}{
  Comparison of \teq{u_{\rm x}(x)/u_1} predicted by the simulation (solid lines)
  with \teq{<\!\! v_{\rm x}\!\!>} (dotted lines) plotted with a linear distance
  scale. The top panel is the example shown in the top left panel of
  Figure~\ref{fig:twoprofiles}, the middle panel is the shock shown in
  Figure~\ref{fig:febprofile}, and the bottom panel is the shock shown in
  Figure~\ref{fig:sfprofile}. While some sharpness of the shock profile is the
  result of our smoothing procedures, the subshock transition is less than
  \teq{1\lambda_0} wide in all cases.         \label{fig:aveflow}
}

\figureout{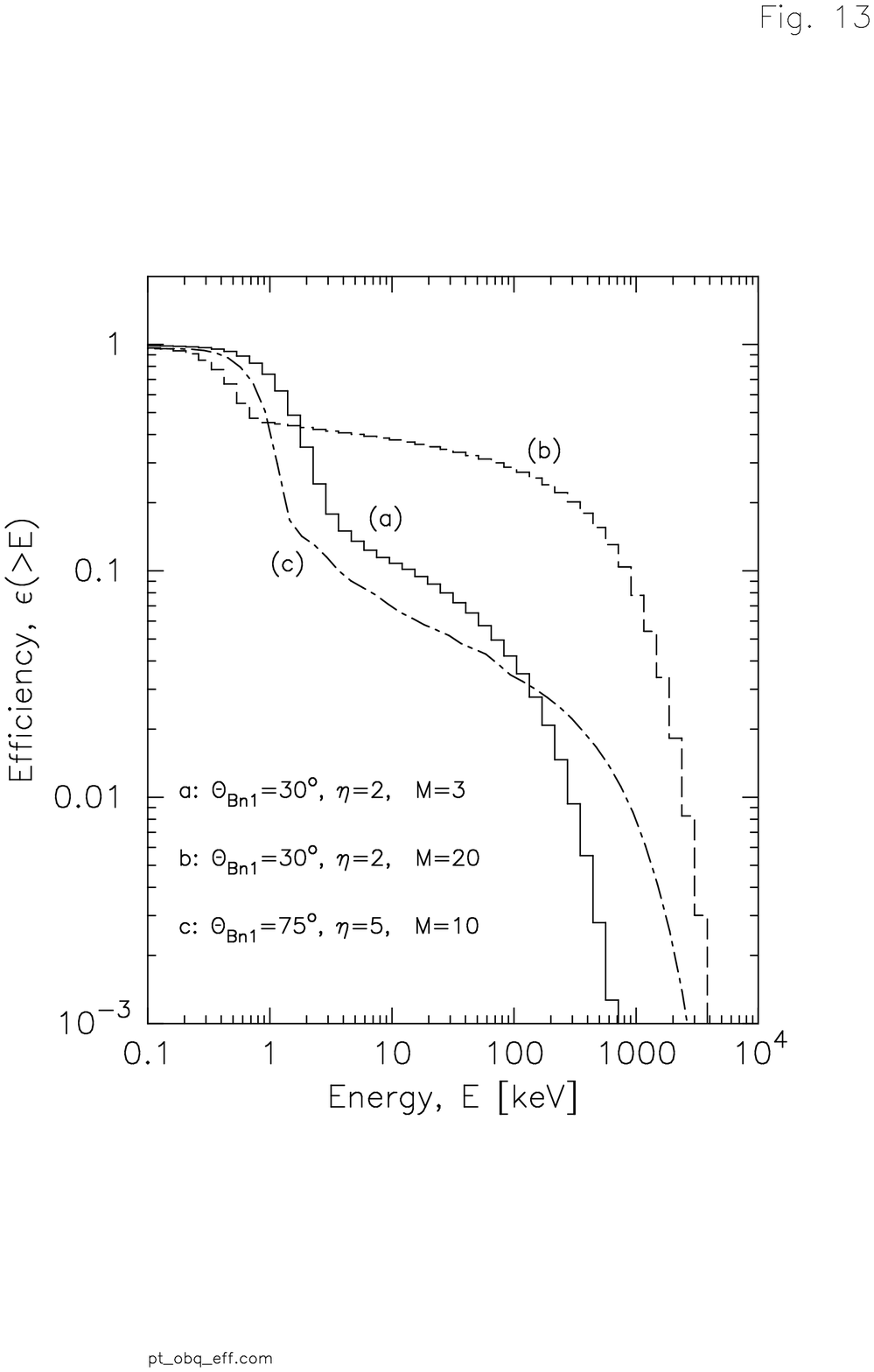}{
  The acceleration efficiency, as defined in equation~(\ref{eq:acceff}), as a
  function of particle energy.  The solid curve (a) is calculated from the LAS
  spectrum shown in Figure~\ref{fig:specone} (\teq{\Tbnone = 30^\circ}, 
  \teq{\eta =2}, \teq{\MA = \MS = 3}), the dashed curve (b) is calculated from
  the smooth shock spectrum shown in Figure~\ref{fig:febspec} (\teq{\Tbnone =
  30^\circ}, \teq{\eta =2}, \teq{\MA = \MS = 20}), and curve (c) is calculated
  from the LAS, smooth shock solution shown in Figure~\ref{fig:sfspec} 
  (\teq{\Tbnone =75^\circ}, \teq{\eta =5}, \teq{\MA = \MS = 10}).  The
  flattening evident at the high energy ends of curves (a) and (b) comes about
  because we have taken \teq{Q_{\rm esc}} to be a constant, neglecting the
  fact that escaping particles leave with a range of energies. 
    \label{fig:figacceff}
}

\figureout{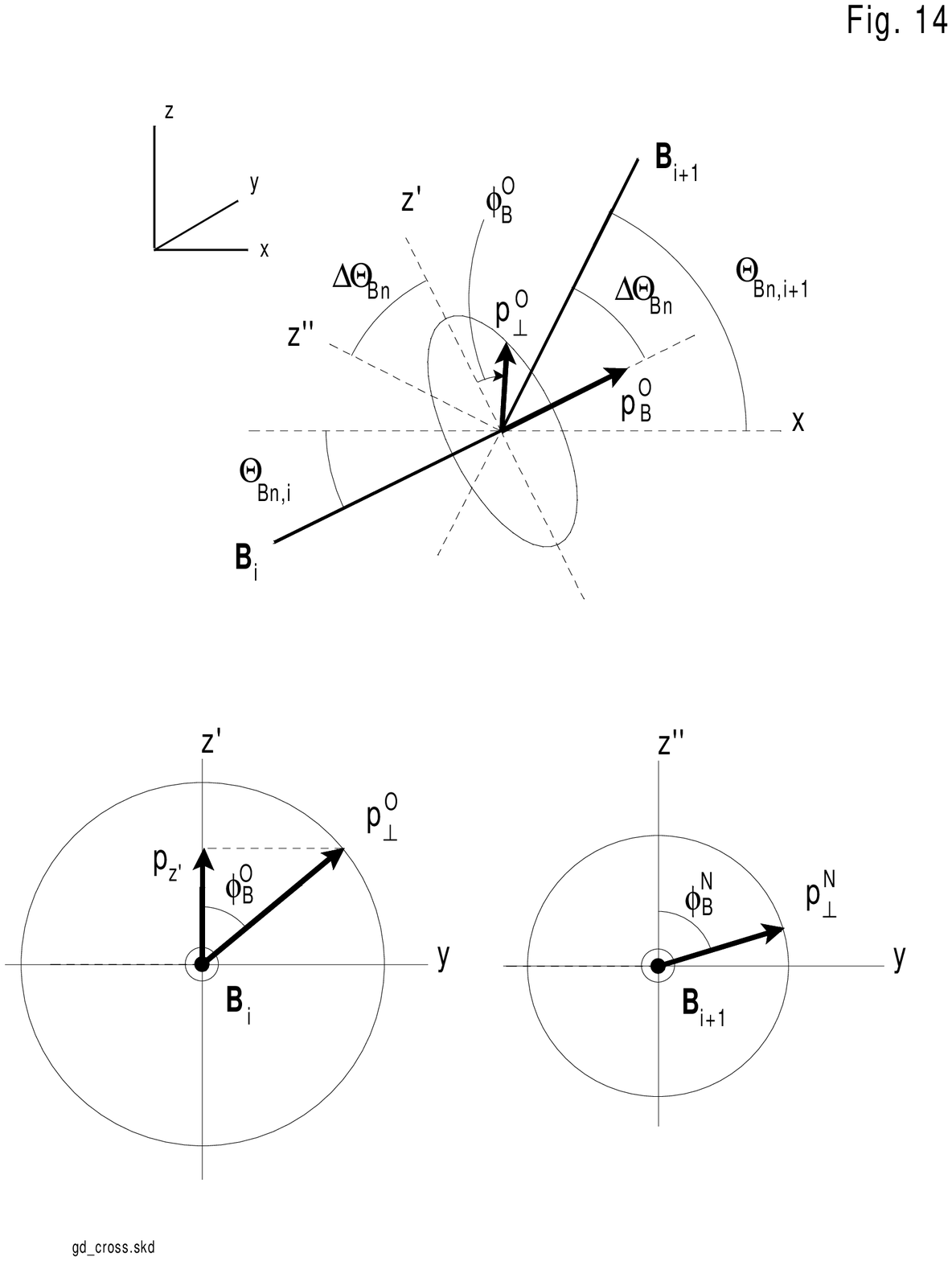}{
  The geometry for calculating the gyroradius and phase of particles
  crossing a grid zone boundary. The boundary occurs at the kink in
  \teq{{\bf B}}.                \label{fig:gdcross}
}                                   

\figureout{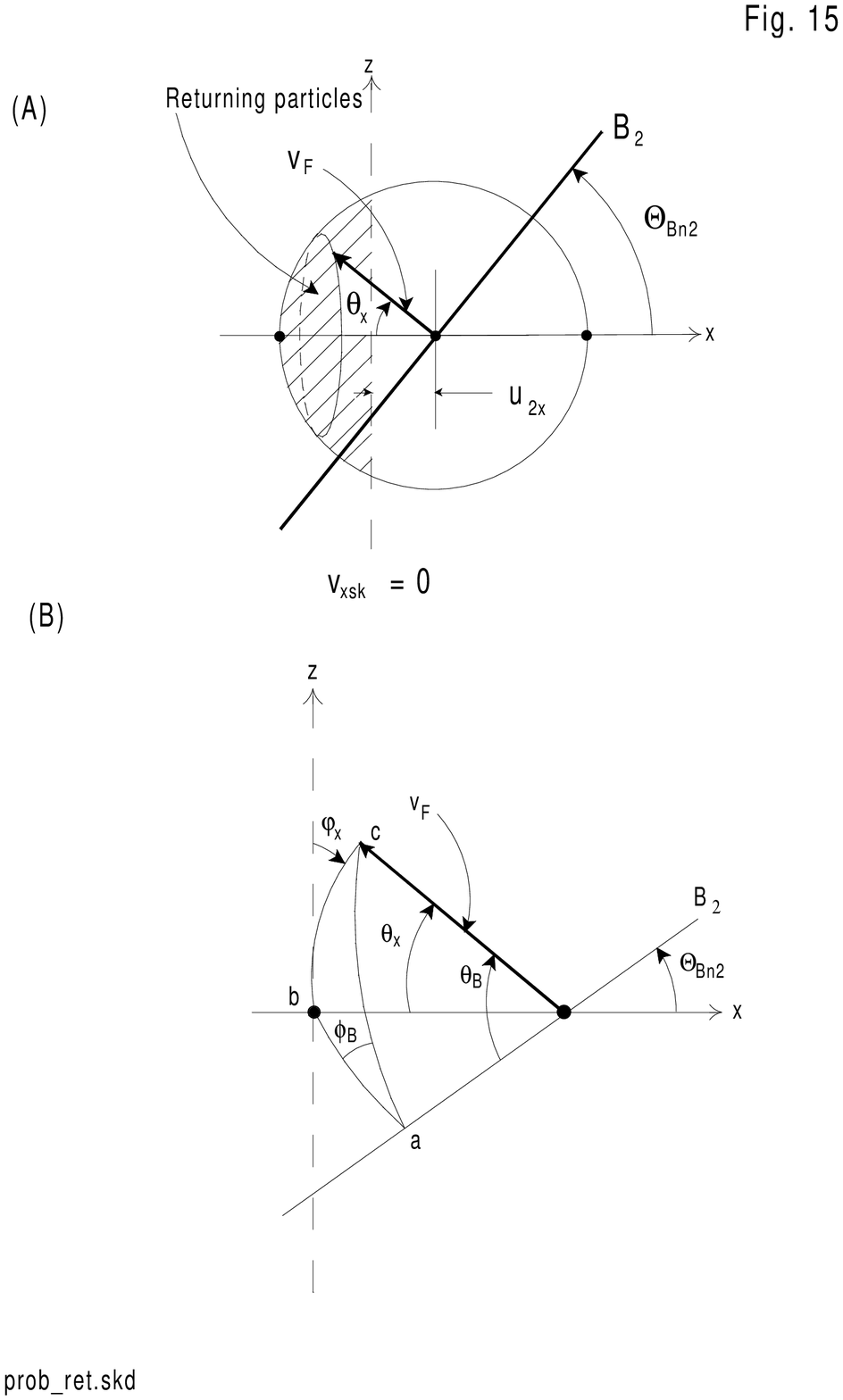}{
  The velocity vectors of returning particles (filled region in figure, A) are
  symmetric about the \teq{x}-axis.  The vertical dashed line is at zero
  velocity in the shock frame and the center of the \teq{\vf} vectors is
  displaced by the flow speed, \teq{u_{\rm x2}}.  Figure B shows the spherical
  triangle, {\it abc}, used to calculate \teq{\PitchB} and \teq{\PhaseB}. 
  Points {\it a} and {\it b} lie in the \teq{x}-\teq{z} plane, while point 
  {\it c} lies off the plane.                   \label{fig:probretone}
}

\clearpage
\psfig{figure=apj96ebjf1.ps,width=6.2in}
\clearpage
\psfig{figure=apj96ebjf2.ps,width=6.2in}
\clearpage
\psfig{figure=apj96ebjf3.ps,width=6.2in}
\clearpage
\psfig{figure=apj96ebjf4.ps,width=6.2in}
\clearpage
\psfig{figure=apj96ebjf5.ps,width=6.2in}
\clearpage
\psfig{figure=apj96ebjf6.ps,width=6.2in}
\clearpage
\psfig{figure=apj96ebjf7.ps,width=6.2in}
\clearpage
\psfig{figure=apj96ebjf8.ps,width=6.2in}
\clearpage
\psfig{figure=apj96ebjf9.ps,width=6.2in}
\clearpage
\psfig{figure=apj96ebjf10.ps,width=6.2in}
\clearpage
\psfig{figure=apj96ebjf11.ps,width=6.2in}
\clearpage
\psfig{figure=apj96ebjf12.ps,width=6.2in}
\clearpage
\psfig{figure=apj96ebjf13.ps,width=6.2in}
\clearpage
\psfig{figure=apj96ebjf14.ps,width=6.2in}
\clearpage
\psfig{figure=apj96ebjf15.ps,width=6.2in}


\begin{references}
%
\reference{}
   Axford, W.~I.: 1965 \planss\vol{13}{115} 
\reference{}
   Axford, W.~I.: 1981 in \it Proc. 17th Int. Cosmic Ray Conf. \rm 
   (Paris), \vol{12}{155} 
\reference{}
   Axford, W.~I. , Leer, E. , and Skadron, G.: 1977 in \it Proc. 15th 
   Int. Cosmic Ray Conf. \rm (Plovdiv), \vol{11}{132}
\reference{}
   Baring, M.~G., Ellison, D.~C., and Jones, F.~C.: 1992 in \it Particle
   Acceleration in Cosmic Plasmas, \rm eds. Zank, G.~P. and Gaisser, T.~K.
   (AIP Conf. Proc. 264; AIP, New York) p.~177
\reference{}
   Baring, M.~G., Ellison, D.~C., and Jones, F.~C.: 1993 \apj\vol{409}{327}
\reference{}
   Baring, M.~G., Ellison, D.~C., and Jones, F.~C.: 1994 \apjs\vol{90}{547}
\reference{}
   Baring, M.~G., Ellison, D.~C., and Jones, F.~C.: 1995 
   \asr\vol{15\, (8/9)}{397}
\reference{}
   Baring, M.~G., Ogilvie, K., Ellison, D.~C., and Forsyth, R.: 1995
   \asr\vol{15\, (8/9)}{388}
\reference{}
   Begelman, M.~C. and Kirk, J. G.: 1990 \apj\vol{353}{66}
\reference{}
   Bell, A.~R.: 1978 \mnras\vol{182}{147}
\reference{}
   Bennett, L. and Ellison, D.~C.: 1995 \jgr\vol{100}{3439}
\reference{} 
   Berezhko, E.~G., Krymsky, G.~F., Ksenofontov, L.~T., and Yelshin, V.~K.: 
   1995 in \it Proc. 24th Int. Cosmic Ray Conf. \rm (Rome), \vol{3}{392}
\reference{} 
   Berezhko, E.~G., Ksenofontov, L.~T., and Yelshin, V.~K.: 1995 
   Nucl. Phys. B (Proc. Suppl.) \vol{39A}{171} 
\reference{} 
   Berezhko, E.~G., Yelshin, V.~K., and Ksenofontov, L.~T.: 1994  
   \app\vol{2}{215}
\reference{}
   Blandford, R.~D. and Eichler, D.: 1987 \physrep\vol{154}{1}
\reference{}
   Blandford, R.~D., and Ostriker, J.~P: 1978 \apjl\vol{221}{L29}
\reference{}
   Boyd, T.~J.~M. and Sanderson, J.~J.: 1969 {\it Plasma Dynamics}, 
   (Barnes and Noble, New York).
\reference{}
   Burgess, D.: 1989 \grl\vol{16}{163}
\reference{}
   Burton, M.~E., Smith, E.~J., Goldstein, B.~E., Balogh, A., Forsyth, R.~J.,
   and Bame, S.~J.: 1992 \grl\vol{19}{1287}
\reference{}
   Decker, R.~B.: 1988 \ssr\vol{48}{195}
\reference{}
   Decker, R.~B., and Vlahos, L.: 1985, \jgr\vol{90}{47}
\reference{}
   de Hoffmann, F. and Teller, E.: 1950 \it Phys. Rev.\vol{80}{692}
\reference{}
   Drury, L.~O'C.: 1983 {\it Rep. Prog. Phys.} \vol{46}{973}
\reference{}
   Eichler, D.: 1984 \apj\vol{277}{429}
\reference{}
   Eichler, D.: 1985 \apj\vol{294}{40}
\reference{}
   Ellison, D.~C.: 1985 \jgr\vol{90}{29}
\reference{}
   Ellison, D.~C., Baring, M.~G. and Jones, F.~C.: 1995 \apj\vol{453}{873}
\reference{}
   Ellison, D.~C., and Eichler, D.: 1984 \apj\vol{286}{691}
\reference{}
   Ellison, D.~C., and Eichler, D.: 1985 \prl\vol{55}{2735}
\reference{}
   Ellison, D.~C., Giacalone, J., Burgess, D., and Schwartz, S.~J.: 1993 
   \jgr\vol{98}{21,085}
\reference{}
   Ellison, D.~C., Jones, F.~C., and Reynolds, S.~P.: 1990 \apj\vol{360}{702}
\reference{}
   Ellison, D.~C., M\"obius, E., and Paschmann, G.: 1990 \apj\vol{352}{376}
\reference{}
   Ellison, D.~C., and Reynolds, S.~P.: 1991 \apj\vol{382}{242}
\reference{}
   Forman, M.~A., Jokipii, J.~R., and Owens, A.~J.: 1974 \apj\vol{192}{535}
\reference{}
   Fulbright, M.~S., and Reynolds, S.~P.: 1990 \apj\vol{357}{591}
\reference{}
   Giacalone, J., Burgess, D., and Schwartz, S.~J.: 1992, in {\it
   Study of the Solar-Terrestrial System,} proc. 26th ESLAB Symposium,
   ESA Special Publication.
\reference{}
   Giacalone, J., Jokipii, J.~R., and K\'ota, J.: 1994 \jgr\vol{99}{19351}
\reference{}
   Gloeckler, G., Geiss, J., Roelof, E.~C., Fisk, L.~A., Ipavich, F.~M., 
   Ogilvie, K.~W., Lanzerotti, L.~J., von Steiger, R., and Wilken, B.:
   1994 \jgr\vol{99}{17,637}
\reference{}
   Hoppe, M.~M., Russell, C.~T., Frank, L.~A., Eastman, T.~E. and
     Greenstadt, E.~W.: 1981 \jgr\vol{86}{4471}
\reference{}
   Ipavich, F.~M., Gloeckler, G., Hamilton, D.~C., Kistler, L.~M.,
   and Gosling, J.~T.: 1988 \grl\vol{15}{1153}
\reference{}
   Jokipii, J.~R.: 1987 \apj\vol{313}{842}
\reference{}
   Jokipii, J.~R.: 1992 \apjl\vol{393}{L41}
\reference{}
   Jokipii, J.~R., and F.~C. Jones, 1996, \prl , submitted.
\reference{}
   Jokipii, J.~R., K\'ota, J., and Giacalone, J.: 1993 \grl\vol{20}{1759}
\reference{}
   Jones, F.~C., and Ellison, D.~C.: 1987 \jgr\vol{92}{11,205}
\reference{}
   Jones, F.~C., and Ellison, D.~C.: 1991 \ssr\vol{58}{259}
\reference{}
   Jones, T.~W., and Kang, H.: 1995 in \it Proc. 24th Int. Cosmic Ray Conf.
   \rm , (Rome) \vol{3}{245}
\reference{}
   Kang, H., and Jones, T.~W.: 1995 \apj\vol{447}{944}
\reference{}
   Kirk, J.~G. and Schneider, P.: 1987 \apj\vol{315}{425}
\reference{}
   Krymsky, G.~F.: 1977 {\it Dokl. Akad. Nauk SSSR}, \vol{243}{1306}
\reference{}
   Liewer, P.~C., Goldstein, B.~E., and  Omidi, N.: 1993 \jgr\vol{98}{15,211}
\reference{}
   Liewer, P.~C., Rath, S., and Goldstein, B.~E.: 1995 \jgr\vol{100}{19,809} 
\reference{}
   Malkov, M.~A., and V\"olk, H.~J.: 1995 in \it Proc. 24th Int. Cosmic 
   Ray Conf. \rm (Rome), \vol{3}{269}
\reference{} 
   Ostrowski, M.: 1988 \aap\vol{206}{169}
\reference{} 
   Ostrowski, M.: 1991 \mnras\vol{249}{551}
\reference{}
   Peacock, J.~A.: 1981 \mnras\vol{196}{135}
\reference{}
   Quest, K.~B.: 1988 \jgr\vol{93}{9649}
\reference{} 
   Takahara F. and Terasawa, T.: 1991 in Proc. ICRR International Symposium: 
     {\it Astrophysical Aspects of the Most Energetic Cosmic Rays}, 
     \rm eds. Nagano, M., Takahara, F. (World Scientific, Singapore) p.~291
\reference{}
   Terasawa, T.: 1979 \it Planet. Space Sci.\vol{27}{193}
\reference{} 
   Toptyghin, I.~N.: 1980 \ssr\vol{26}{157}
\reference{}
   V\"olk, H.~J.: 1984, in {\it High Energy Astrophysics,
Proc. 19th Rencontre de Moriond,}
ed. Tran Than Van (Gif-sur-Yvette: Editions Fronti\`eres), p. 281.

\reference{}
   Winske, D., Omidi, N., Quest, K.~B. and Thomas, V.~A.: 1990
     \jgr\vol{95}{18,821}
\reference{}
   Zachary, A.: 1987 {\it Resonant Alfv\'en Wave Instabilities
   Driven by Streaming Fast Particles,} Ph.D. thesis, Lawrence Livermore
   National Laboratory, University of California.
\reference{}
   Zank, G.~P., Webb, G.~M., and Donohue, D.~J.: 1993 \apj\vol{406}{67}

\end{references}
\end{document}